\documentclass[12pt]{article}
\usepackage[T1]{fontenc}
\usepackage[utf8]{inputenc}
\usepackage[height=8.85in,width=6.45in]{geometry}
\usepackage{lmodern}

\usepackage{amsmath}
\usepackage{amssymb}
\usepackage{mathtools}
\numberwithin{equation}{section}
\usepackage{slashed}
\usepackage{braket}
\usepackage[svgnames]{xcolor}
\usepackage[colorlinks,citecolor=DarkGreen,linkcolor=FireBrick,urlcolor=FireBrick,linktocpage,breaklinks=true]{hyperref}
\usepackage{cite}
\usepackage{graphicx}
\usepackage{multirow}
\usepackage{times}
\usepackage{courier}
\usepackage{bm}
\usepackage{subfig}
\usepackage{xcolor}
\usepackage{mdframed}
\usepackage{mathrsfs}
\usepackage{cleveref}
\usepackage{tikz-cd}
\usetikzlibrary{decorations.markings}

\crefformat{section}{#2Sec.~#1#3}
\crefformat{appendix}{#2App.~#1#3}
\crefformat{table}{#2Table~#1#3}

\newenvironment{claim}{\begin{mdframed}[linecolor=black!0,backgroundcolor=black!4]\noindent\itshape\ignorespaces}{\end{mdframed}}

\renewenvironment{figure}[1][]{
  \begin{originalfigure}[#1]
    \begin{mdframed}[linecolor=black!0,backgroundcolor=black!4]
}{
    \end{mdframed}
  \end{originalfigure}
}

\renewenvironment{table}[1][]{
  \begin{originaltable}[#1]
    \begin{mdframed}[linecolor=black!0,backgroundcolor=black!4]
}{
    \end{mdframed}
  \end{originaltable}
}

\newcommand{\bC}{\mathbb{C}}
\newcommand{\bZ}{\mathbb{Z}}
\newcommand{\bR}{\mathbb{R}}

\newcommand{\cH}{\mathcal{H}}

\newcommand{\Spin}{\mathrm{Spin}}
\newcommand{\SO}{\mathrm{SO}}

\newcommand{\ii}{\mathrm{i}}
\newcommand{\ab}{\mathrm{ab}}

\DeclareMathOperator{\Hom}{Hom}
\newcommand{\Arf}{\mathrm{Arf}}

\newcommand{\pt}{\mathrm{pt}}

\begin{document}

\begin{titlepage}

\begin{flushright}
\end{flushright}

\vskip 3cm

\begin{center}

{\Large \bfseries On Generalised Discrete Torsion}

\vskip 2cm

Philip Boyle Smith$^{1,2}$ and
Yuji Tachikawa$^3$
\vskip 1cm

\begin{tabular}{ll}

1 & SISSA, via Bonomea 265, 34136 Trieste, Italy \\
2 & INFN, Sezione di Trieste, via Valerio 2, 34127 Trieste, Italy \\
3 & Kavli Institute for the Physics and Mathematics of the Universe (WPI), \\
& University of Tokyo, Kashiwa, Chiba 277-8583, Japan

\end{tabular}

\vskip 2cm

\end{center}

\noindent
For a 2d gauged sigma model with target space $M$ and discrete gauge group $G$,
we consider a generalisation of Vafa's discrete torsion $H^2(BG; U(1))$
that assigns different local discrete torsion phases to different singular loci of the orbifold $M/G$.
Our generalised discrete torsion lives in $H^2_G(M; U(1))$,
and gives a consistent implementation of Gaberdiel and Kaste's prescription
for inserting such local discrete torsion phases by hand at higher genus.
We revisit the original application to $T^6/\bZ_2^2$ and $T^7/\bZ_2^3$ orbifold CFTs,
and determine what smooth Calabi--Yau and $G_2$ geometries result
from different choices of the generalised discrete torsion.
We find that the local discrete torsion phases can be different from each other,
but are not completely independent either;
in the $T^7/\bZ_2^3$ case for example,
the orbifold CFTs only realise $3$ out of the $9$ possible Betti numbers of $G_2$ resolutions constructed by Joyce.

\end{titlepage}

\setcounter{tocdepth}{2}
\tableofcontents

\section{Introduction}

Discrete torsion is an additional phase we can introduce
when we take an orbifold of a two-dimensional quantum field theory
by a finite group $G$, depending on the choice of the twists by $G$
on the worldsheet \cite{Vafa:1986wx}.
As already explained there, this phase is controlled by a cohomology class $\alpha\in H^2(BG;U(1))$.
One of the simplest cases is to take $G=\bZ_2\times \bZ_2$,
for which $H^2(B(\bZ_2\times\bZ_2);U(1))=\bZ_2$. Then we have two variants of the $\bZ_2\times \bZ_2$ orbifold.

Furthermore, when $G$ acts on a sigma model with target space $M$,
discrete torsion is known to affect the structure of the orbifold singularity of
the quotient space $M/G$ \cite{Vafa:1994rv}.
Consider the simple case of the Calabi--Yau orbifold
\begin{equation} \label{local-sing}
    ((\bC^2 / \bZ_2) \times T^2) / \bZ_2,
\end{equation}
treated in that reference.
To specify the orbifold action,
let $u,v$ denote the two coordinates of $\bC^2$, and $s$ denote the complex coordinate of $T^2$ with periodic identification.
Then the $\bZ_2$ inside acts by flipping the signs of $u$ and $v$,
and the $\bZ_2$ outside acts by flipping those of $u$ and $s$.
This geometry has a singular locus extending along the $T^2/\bZ_2$ direction,
and can be made smooth in two ways, either by `resolving' by inserting a small 2-cycle,
contributing by $+1$ to the Hodge number $h^{1,1}$,
or by `deforming' by inserting a small 3-cycle, contributing by $+1$ to the Hodge number $h^{2,1}$.\footnote{%
One way to see it is as follows. Regard $\bC^2/\bZ_2$
as the Kaluza--Klein monopole situated at the origin $0\in \bR^3$ of charge $2$.
A three-parameter deformation is provided
by the configuration of two Kaluza--Klein monopoles of charge $+1$
at positions $\pm (x_0,y_0,z_0) \in \bR^3$.
This creates a finite-sized $S^2$ in the geometry.
Now, the action $(u,v)\mapsto (-u,v)$ translates to the action $(x,y,z)\mapsto (x,-y,-z)$ of $\bR^3$.
Then, the parameter $x_0$ is even while $y_0$, $z_0$ are odd
under the second $\bZ_2$ action.

We now need to fibre these deformations over $T^2$, in a way compatible with the second $\bZ_2$ action.
The $\bZ_2$-even deformation can be simply set to be constant, creating an $S^2$ and
$S^2\times (T^2/\bZ_2)$ in the geometry.
The two $\bZ_2$-odd deformations, in contrast, need to vanish at the $\bZ_2$ fixed points in $T^2$,
creating two $S^3$s in the geometry.
Conventionally, the introduction of $S^2$ is called the resolution,
while the introduction of $S^3$ is called the deformation.
}

Let us realise the sigma model on this orbifold by first taking the flat sigma model on $\bC^2\times T^2$
and gauge the discrete symmetry group $\bZ_2\times \bZ_2$.
The resulting two-dimensional theory is not singular,
and the Hodge numbers $h^{p,q}$ can be computed without any issue.
It turns out that
there is a $+1$ contribution to $h^{1,1}$ or to $h^{2,1}$
depending on whether or not we include the non-trivial discrete torsion for the orbifold group $\bZ_2\times \bZ_2$.
In a certain sense, we can say that the discrete gauging of the flat sigma model
produces the orbifold geometry which is very slightly `resolved' or `deformed', depending on the value of the discrete torsion.

This analysis can be readily generalised to the case of the compact orbifold
\begin{equation}
    (T^2 \times T^2 \times T^2) / (\bZ_2 \times \bZ_2).
\end{equation}
This has $48$ singular loci of the form \eqref{local-sing}, which in total provides a $+48$ contribution to
$h^{1,1}$ or to $h^{2,1}$, again depending on the value of the discrete torsion.
But this begs a natural question: is it possible to choose whether we `resolve' or `deform' the singularity,
depending on the singular loci?
At the level of classical geometry, it was demonstrated mathematically that we can actually
make such choices, not just in the context of six-dimensional Calabi--Yau manifolds
but also for seven-dimensional $G_2$ manifolds and eight-dimensional $Spin(7)$ manifolds,
as first found by Joyce in 1996 and summarized later in his textbook \cite{JoyceBook}.
Then one wonders if it is possible to assign discrete torsion phases independently
for each singular locus, to mimic this choice in classical geometry, as already asked in \cite{Joyce:1998en}.
Gaberdiel and Kaste considered this question in \cite{Gaberdiel:2004vx},
and suggested that such independent assignments were possible,
but the consistency at higher genus was unclear.

In this paper, we provide a consistently generalised version of discrete torsion,
now taking values not in the group cohomology $H^2(BG;U(1))$
but in the equivariant cohomology $H^2_G(M;U(1))$ of the target space.
We find that the discrete torsion phases
at the singular loci are no longer forced to be all the same,
but are not completely independent from each other, either.
We will investigate this interesting phenomenon by a detailed study of
the Calabi--Yau orbifold $T^6/\bZ_2^2$ given above as the first example,
and the $G_2$ orbifold $T^7/\bZ_2^3$ as the second.

The fact that the phase controlled by the equivariant cohomology $H^2_G(M;U(1))$
can be included in the discrete gauging
has been known for some time.
Indeed, already in the original paper \cite{Vafa:1986wx}, it was mentioned that discrete torsion should be considered as part of the $B$-field on the orbifold geometry,
and this point was further carefully examined using the language of gerbes by a series of papers by Sharpe,
see e.g.~\cite{Sharpe:2000ki}.
From this description it should mathematically follow that flat $B$-fields on an orbifold $M/G$ are parameterised by
the equivariant cohomology given above.
In particular, all the generalised discrete torsion phases we use in this paper,
in the case of torus orbifolds $T^n/G$, were already explicitly discussed
in another paper by Sharpe \cite{Sharpe:2003cs}.
The use of equivariant cohomology to classify topological terms in gauged sigma models was also discussed in \cite{Davighi:2020vcm}.
The only new point in this paper, then, is the realisation that
the inclusion of these known effects provides a consistent version of the construction of Gaberdiel and Kaste.

The rest of the paper is organised as follows. In \cref{sec:generalities} we motivate equivariant cohomology as a notion of generalised discrete torsion. In \cref{sec:t6} and \cref{sec:t7} we consider the effects of such generalised discrete torsion on the Calabi--Yau orbifold $T^6 / \bZ_2^2$ and the $G_2$ orbifold $T^7 / \bZ_2^3$, by computing the dependence of the Hodge and Betti numbers respectively on the discrete torsion class. We conclude with two appendices: in \cref{app:sharpe} we connect to the formalism of~\cite{Sharpe:2003cs}, while in \cref{app:hcalc} we collect some computations in group cohomology.

\section{Generalities} \label{sec:generalities}

Let us start by studying the general structure of various phases we can introduce in the formulation of two-dimensional theories.

\subsection{Ordinary discrete torsion} \label{sec:odt}

The original discrete torsion of \cite{Vafa:1986wx} is a phase
we can add when we gauge a discrete group $G$.
Let us remind ourselves how it is given by a cohomology class $\omega\in H^2(BG;U(1))$.
For ease of reading, we use an additive notation for $U(1)$, by regarding it as $\bR/\bZ$.

Let $\Sigma$ be the worldsheet. A background $G$-gauge field can then be represented by
a map $f:\Sigma\to BG$ where $BG$ is the classifying space of $G$.
We pick a cohomology class $\omega\in H^2(BG;U(1))$, and pull it back via $f$
to obtain $f^*(\omega)\in H^2(\Sigma;U(1))$.
The phase is then the pairing $\int_\Sigma f^*(\omega) \in U(1)$,
which enters the exponentiated Euclidean action as $\exp(2\pi\ii \int_\Sigma f^*(\omega))$.

An algebraic definition of $H^2(BG;U(1))$ is given by
considering 2-cocycles, which are functions $\omega:G\times G\to U(1)$
satisfying
\begin{equation} \label{cocycle}
    \omega(g,h) + \omega(gh,k) = \omega(g,hk) + \omega(h,k)
\end{equation}
and then identifying two 2-cocycles $\omega(g,h)$ and $\tilde\omega(g,h)$ related by
\begin{equation} \label{coboundary}
    \tilde\omega(g,h) = \omega(g,h) + \psi(g) + \psi(h) - \psi(gh)
\end{equation}
for a 1-cochain $\psi : G \to U(1)$.

Then, for a torus $T^2$ with commuting twists $g,h$ along its two 1-cycles,
the associated phase is given by
\begin{equation} \label{pairing}
    \langle g,h \rangle_\omega \coloneqq \int_{T^2} f^*(\omega) = \omega(g,h) - \omega(h,g) \in U(1),
\end{equation}
where we defined the bracket notation for later use.
We will often drop the subscript $\omega$, if there would be no source of confusion.

Clearly, we have $\langle g,h \rangle_\omega = \langle g,h \rangle_{\tilde\omega}$
when $\omega$ and $\tilde \omega$ are related as in \eqref{coboundary}.
Therefore the pairing is determined by the cohomology class $\omega\in H^2(BG;U(1))$.
From the expression \eqref{pairing},
we have (i) $\langle g,h\rangle_\omega=-\langle h,g\rangle_\omega$
and (ii) $\langle g,g\rangle_\omega=0$.
Furthermore, we can show that (iii) this pairing $\langle g,h\rangle_\omega$ is
a homomorphism for both variables $g,h$.

If $G$ is finitely generated and abelian,
$U(1)$-valued pairings on $G$ satisfying the three conditions (i), (ii) and (iii)
are in one-to-one correspondence with elements of $H^2(BG;U(1))$.
In such cases, the function $\varphi^{\omega}_g: G\to U(1)$ determined by $\varphi^\omega_g(h)=\langle g,h\rangle_\omega$
is a one-dimensional representation of $G$, i.e.~specifies a charge of $G$.
In the Hamiltonian language, this means that the inclusion of a discrete torsion phase specified by $\omega$
shifts the charge of $G$ in the $g$-twisted sector by $\varphi^{\omega}_g$.

Take $G=\bZ_2\times \bZ_2$ as an example. Let $\alpha$ and $\beta$ be the generators of two factors of $\bZ_2$.
Then a pairing satisfying the conditions (i), (ii) and (iii) is specified by the value
\begin{equation}
    \langle \alpha,\beta \rangle \in \{0,\tfrac12\} \subseteq U(1).
\end{equation}
This means that $H^2(B(\bZ_2\times \bZ_2);U(1))=\bZ_2$,
and a nontrivial discrete torsion changes the charge under $\beta$ of the $\alpha$-twisted sector, and so on.

\subsection{Ordinary $B$-field} \label{sec:B}

Let us next discuss the $B$-field.
It is useful to start by comparing against the properties of the $U(1)$ gauge field.

A $U(1)$ gauge field is locally described by a one-form $A$, defined up to gauge transformations.
The gauge-invariant field strength is a globally well-defined two-form $F$.
Mathematically, it is specified by a $U(1)$ connection.
A flat $U(1)$ gauge field, i.e.~one with $F=0$, is specified by its holonomy $\int_C A\in U(1)$
for 1-cycles $C$.
This means that a flat $U(1)$ gauge field on $M$ is specified by a homomorphism $H_1(M;\bZ)\to U(1)$,
or equivalently an element in $H^1(M;U(1))$.

A $B$-field is a two-form version of the abelian gauge field.
Locally it is given by a two-form $B$, whose gauge invariant field strength $H$ is a globally well-defined three-form $H$.
A flat $B$-field, i.e.~one with $H=0$, is specified by its holonomy $\int_\Sigma B\in U(1)$
for 2-cycles $B$.
Then a flat $B$-field on $M$ is specified by a homomorphism $H_2(M;\bZ)\to U(1)$,
or equivalently an element in $H^2(M;U(1))$.
A mathematical formulation of the $B$-field including non-flat ones is provided by the concept of the gerbe,
see e.g.~\cite{Sharpe:1999pv,Sharpe:1999xw}.

In string theory, more general abelian gauge fields locally given by $p$-forms with $p>2$ also appear.
The definition of $U(1)$ connections for $p=1$ and gerbes for $p=2$ does not directly generalise
to the cases with higher $p$.
Instead it is more common to use differential cohomology classes $\hat H^{p+1}(M)$,
whose formulation applies uniformly for all $p$, to describe these form-fields.
In this paper, we only treat flat $B$-fields, for which we can simply use $H^2(M;U(1))$ as its mathematical description, without invoking anything fancier.

The flat $B$-field specified by $\omega\in H^2(M;U(1))$ affects the sigma model action in the following manner.
Let $f: \Sigma\to M$ be the sigma model map. We can pull back the class $\omega$ via this map $f$,
and obtain the class $f^*(\omega)\in H^2(\Sigma;U(1))$.
Then we have a contribution to the exponentiated Euclidean action given by $\exp(2\pi\ii \int_\Sigma f^*(\omega))$.
By comparing this discussion with the one we gave in \cref{sec:odt},
we see that it is exactly the same, except that we now replaced $BG$ by $M$.

\subsection{Generalised discrete torsion}

Let us now consider the combined situation,
where we consider a sigma model $M$ with an isometry $G$, which we also gauge.
How should we formulate the data of a flat $B$-field?

When the $G$-bundle over the worldsheet $\Sigma$ is nontrivial,
the sigma model field is not a map to $M$.
Rather, we have an $M$-bundle $M\to X\to \Sigma$ whose fibre bundle structure is determined by
the given $G$-bundle and the $G$-action on $M$,
and the sigma model field is a section of this $M$-bundle.
More abstractly, we have a universal $G$-bundle $G\to EG\to BG$,
with which we construct a universal $M$-bundle $M\to (EG \times M)/G \to BG$,
where the quotient in the middle term is via the diagonal action of $G$ on $EG$ and $M$.
Then, the $G$-gauge bundle together with the sigma model field on the worldsheet $\Sigma$
defines a map $f:\Sigma\to (EG\times M)/G$.
Now, we can pick any cohomology class $\omega \in H^2((EG\times M)/G;U(1))$,
pull it back via $f$ to obtain $f^*(\omega)\in H^2(\Sigma;U(1))$,
and use the resulting $\int_\Sigma f^*(\omega)$ in the sigma model action, exactly as before.

Mathematicians have a special notation for the resulting cohomology group, which is
\begin{equation}
    H^p_G(M;A) \coloneqq H^p((EG\times M)/G;A)
\end{equation}
for a space $M$ with a $G$-action,
where we wrote down a general version for arbitrary degree $p$ and the coefficient abelian group $A$.
The resulting object is called the $G$-equivariant cohomology group of $M$,
and the formation of the space $(EG\times M)/G$ from $G$ and $M$ is known as the Borel construction.

Note that this construction reduces to the group $H^2(BG;U(1))$ when $M$ is a point, since $EG/G=BG$
and therefore $H^2_G(\pt;U(1))=H^2(BG;U(1))$.
This reproduces the ordinary discrete torsion we discussed in \cref{sec:odt}.
Note also that, more obviously, this construction reduces to the group $H^2(M;U(1))$ when $G=1$,
and therefore reproduces the flat $B$-field we discussed in \cref{sec:B}.
This means that the construction here is a natural amalgamation of the two effects discussed in \cref{sec:odt} and \cref{sec:B}.

According to the modern understanding of SPT phases \cite{Freed:2016rqq} in terms of cobordism, we should strictly replace $H^p(\frac{EG \times M}{G}; U(1))$ with its bordism refinement $\Hom(\widetilde{\Omega}_\SO^p(\frac{EG \times M}{G}, U(1))$. However when $p = 2$, the two coincide, so we can ignore this subtlety. Furthermore, when the worldsheet includes fermions, we should consider the further refinement $\Hom(\widetilde{\Omega}_\Spin^2(\frac{EG \times M}{G}, U(1))$ to spin bordism. This group is $H^2(\frac{EG \times M}{G}; U(1)) \times H^1(\frac{EG \times M}{G}; \bZ_2)$, where the first term is the discrete torsion we have discussed so far, and the new term corresponds to `fermionic discrete torsion' $(-1)^{\Arf[\rho + f^*(A)] + \Arf[\rho]}$ where $A \in H^1(\frac{EG \times M}{G}; \bZ_2)$. Such discrete torsion makes the orbifold CFT non-geometric (with vanishing zeroth Betti number) and breaks spacetime SUSY, so we do not consider it further here. Thus for our purposes, $H^2(\frac{EG \times M}{G}; U(1))$ shall suffice.

\subsubsection*{The continuous part} \label{sec:contpart}

A subgroup of the equivariant cohomology group $H^2_G(M; U(1))$ simply describes the part of the $B$-field on $M$ that survives after gauging $G$.
Let $\omega \in H^2(M; \bR)^G$ be such a $B$-field, represented by a closed, $G$-invariant differential 2-form on $M$.
Then we can write down the topological term $\exp(2\pi\ii \int_\Sigma f^*(\omega))$ in the action, establishing a natural map
\begin{equation}
    H^2(M; \bR)^G \to H^2_G(M; U(1)).
\end{equation}
This map can be shown to surject onto the connected component of the identity, which furthermore takes the form $U(1)^{\dim(H^2(M; \bR)^G)}$. Hence the identity component $H^2_G(M; U(1))_0$ is precisely the remnant of the $B$-field on $M$.

Because the continuous part of the $B$-field only affects the wrapped states of the orbifold CFT,
it does not affect the low-energy physics,
including how each singularity is resolved or deformed.
Said mathematically, the $B$-field is a continuous parameter, so discrete data such as the topology of the resolved space cannot depend on it. This leads us to consider the group of deformation classes
\begin{equation} \label{eq:hdefclasses}
    [H^2_G(M; U(1))] = H^2_G(M; U(1)) / H^2_G(M; U(1))_0
\end{equation}
given by ignoring the $B$-fields. This is a finite, abelian group, and captures the part of $H^2_G(M; U(1))$ that truly corresponds to generalised discrete torsion.

\subsubsection*{Local discrete torsion} \label{sec:loc}

As promised, we can recast \eqref{eq:hdefclasses} as a local version of ordinary discrete torsion.
Take a subset $N \subseteq M$ such that a subgroup $H \leq G$ maps $N$ to $N$.
Then since $EG$ is also a space satisfying the requirements of $EH$,
we can define the $H$-equivariant cohomology of $N$ as $H^2((EG \times N) / H; U(1))$.
This lets us define the pullback $\iota_N^*: H_G^2(M;U(1)) \to H_H^2(N;U(1))$ as in the case of ordinary cohomology.
When $N$ corresponds to a singular locus of the classical orbifold $M/G$, we will see that the phase $\iota_N^*(\omega)$ controls how $N$ is resolved in the orbifold CFT, or whether it is resolved at all.

As a particular case, when a point $x\in M$ is fixed by $H \leq G$, the inclusion $\iota_x : \{x\} \hookrightarrow M$
can be used to define $\iota_x^*(\omega)\in H_H^2(\{x\};U(1))=H^2(BH;U(1))$
from $\omega\in H^2_G(M;U(1))$.
This object $\iota_x^*(\omega)$ can be thought of as the local value of the discrete torsion
at a $H$-fixed point in $M$.

\subsubsection*{Torus orbifolds}

Next, consider the special case of $M=T^n$ acted on by a finite group $G$.
In this case, $T^n$ is actually $\bR^n/\bZ^n$,
and the entire geometry is specified by the extension $1 \to \bZ^n \to \hat{G} \to G \to 1$.
In other words, we have the equality
\begin{equation}
    T^n / G = \bR^n / \hat{G}.
\end{equation}
In this particular case, as $\bR^n$ is topologically trivial,
we have the equality
\begin{equation} \label{eq:ghat}
    H^2_G(T^n;U(1)) = H^2(B\hat{G}; U(1)).
\end{equation}
Therefore, the generalised discrete torsion specified by the equivariant cohomology group
in this particular case
is simply the ordinary discrete torsion associated to the discrete but infinite group $\hat{G}$.
Note that when $G$ is trivial, the elements of
$H^2(T^n;U(1))=H^2(B\bZ^n;U(1))$ are specified by phases $\langle e_i, e_j \rangle \in U(1)$,
where $e_i$ is the generator of the $i$-th $\bZ$.
We can identify this quantity $\langle e_i, e_j \rangle$
with the integral $\int_{T^2_{ij}} B$ of the $B$-field over $T^2_{ij}$, the torus $T^2 \subseteq T^n$
in the $i$-th and $j$-th directions.

\subsubsection*{Discrete torsion from gauging} \label{sec:gauging}

In the actual cases we consider, $G$ is abelian. Suppose the $G$-action on $\bZ^n$ preserves a sublattice $\Gamma$. In such a case, we can regard the original $T^n$ as a quotient
\begin{equation}
    T^n = \widetilde{T}^n / (\bZ^n / \Gamma)
\end{equation}
of a larger space $\widetilde{T}^n = \bR^n / \Gamma$, with a compatible action of the larger group $\hat{G} / \Gamma$. Then
\begin{equation}
    T^n / G = \widetilde{T}^n / (\hat{G} / \Gamma)
\end{equation}
meaning that some of the generalised discrete torsion phases
can be obtained by simply considering the ordinary discrete torsion phases of $\hat{G} / \Gamma$,
by considering the original torus as the $\bZ^n / \Gamma$ orbifold of a larger torus. For example, choosing $\Gamma = \{0\}$ recovers the most general possible discrete torsion phase, but requires us to gauge the large and potentially complicated non-abelian group $\hat{G}$. At the other extreme, choosing $\Gamma = \bZ^n$ only recovers the ordinary discrete torsion of $G$, but the gauging is by the original and simpler abelian group $G$.

For ease of calculation, it is in our interest to find a middle ground where we recover as many discrete torsion phases as possible, but the gauging is still easy to do, that is, we want to minimise $\Gamma$ while keeping $\hat{G} / \Gamma$ abelian. The answer to this problem is to take $\Gamma$ to be the derived subgroup $\hat{G}'=[\hat{G},\hat{G}]$, whereupon $\hat{G} / \Gamma$ becomes the abelianisation $\hat{G}^\ab$. Thus a subset of discrete torsion phases will come from the natural map
\begin{equation} \label{eq:gabnaturalmap}
    H^2(\hat{G}^\ab; U(1)) \to H^2(\hat{G}; U(1)) = H^2_G(T^n; U(1))
\end{equation}
by regarding $T^n/G$ as $\widetilde{T}^n/\hat{G}^\ab$. It should be noted, however, that there is no guarantee that this map is either surjective or injective.
This still provides an easy way to produce examples of generalised discrete torsion phases
more general than discrete torsion phases of $G$ itself.

In the examples we consider, the natural map \eqref{eq:gabnaturalmap} actually is surjective onto the group of deformation classes \eqref{eq:hdefclasses}. So this simple construction will suffice for all our needs.

\subsubsection*{Relation to gerbes and differential cohomology classes}

Finally, it is to be noted that flat $B$-fields on $T^n/\Gamma$ and much more were analysed by Sharpe in \cite{Sharpe:2003cs}
using a formulation in terms of gerbes.
We explicitly demonstrate in \cref{app:sharpe} that the effect described there by Sharpe,
in the case of geometric orbifolds of $T^n$,
is equal to the one described by \eqref{eq:ghat}.
In \cite{Davighi:2020vcm}, equivariant differential cohomology classes were used instead,
to describe topological terms in gauge theories, which include orbifolds by finite groups.
In this case, it is clear that the flat part is the same as the ordinary equivariant cohomology.
Furthermore, in \cite{ParkRedden}, equivariant differential cohomology classes
of degree three were shown to represent the same objects as equivariant gerbes. 
Therefore, independent of the formulation used,
it is a general fact that the flat part of the $B$-field on a $G$-orbifold of $M$ is described by the equivariant cohomology group $H^2_G(M;U(1))$.

\section{$T^6/\bZ_2^2$} \label{sec:t6}

In this section we consider a Calabi--Yau orbifold $T^6/\bZ_2^2$, considered by \cite{Vafa:1994rv,JoyceBook,Gaberdiel:2004vx}.

\subsection{Geometry} \label{sec:t6geom}

Let $x_1,\dots,x_6$ denote the six periodic coordinates of $T^6$, with $x_i \sim x_i + 1$. We then let the two generators $\alpha$ and $\beta$ of $\bZ_2^2$ act by the signs
\begin{equation} \label{eq:t6signs}
    \begin{array}{c|cccccc}
        & x_1 & x_2 & x_3 & x_4 & x_5 & x_6 \\
        \hline
        \alpha & +1 & +1 & -1 & -1 & -1 & -1 \\
        \beta & -1 & -1 & +1 & +1 & -1 & -1
    \end{array}
\end{equation}
The $\alpha$-fixed locus consists of $16$ copies of $T^2$ localised at $(x_3,x_4,x_5,x_6)=\frac{1}{2}(a_3,a_4,a_5,a_6)$ with $a_{3,4,5,6} \in \{0,1\}$. The $\beta$- and $\alpha\beta$-fixed loci are also $16$ copies of $T^2$, differing only in the indices. Each fixed $T^2$ is mapped into itself under the whole $\bZ_2^2$. The $\alpha$-, $\beta$-, and $\alpha\beta$-fixed loci intersect at the $64$ fixed points $(x_1,\dots,x_6)=\frac{1}{2}(a_1,\dots,a_6)$ with $a_{1,\dots,6} \in \{0,1\}$.

\subsection{Cohomology} \label{sec:t6coh}

To access a subset of the generalised discrete torsion phases for $T^6 / \bZ_2^2$, we will write $T^6 = \widetilde{T}^6 / \bZ_2^6$ by doubling the radii of all $6$ circles and gauging the $6$ half-shifts $S_i : \tilde{x}_i \mapsto \tilde{x}_i + \tfrac{1}{2}$. The orbifold $T^6/\bZ_2^2$ can then be rewritten as $\widetilde{T}^6/\bZ_2^8$ where the $\bZ_2^8$ symmetry is generated by $\{S_1, \dots, S_6, \alpha, \beta\}$.%
\footnote{\label{foot:T6hatG}%
According to the general discussion in \cref{sec:gauging}, this is the unique rewriting of the orbifold involving the largest possible abelian group. Indeed, the group $\hat{G}$ such that $T^6/\bZ_2^2 = \bR^6/\hat{G}$ takes the form of a semidirect product $\bZ^6 \rtimes \bZ_2^2$, where the generators $T_{1,\dots,6}$ of the $\bZ^6$ factor are acted on by the generators $\alpha,\beta$ of the $\bZ_2^2$ factor as \[\alpha T_i \alpha^{-1} = T_i^{({+1}, {+1}, {-1}, {-1}, {-1}, {-1})_i} \qquad \beta T_i \beta^{-1} = T_i^{({-1}, {-1}, {+1}, {+1}, {-1}, {-1})_i}\] One then computes that the derived subgroup $\hat{G}'$ is generated by $T_1^2, \dots, T_6^2$, and therefore that $\hat{G}^\ab = \bZ_2^8$ generated by $T_1, \dots, T_6, \alpha, \beta$. On $\widetilde{T}^6 = \mathbb{R}^6/\hat{G}'$, these generators act as $S_1, \dots, S_6, \alpha, \beta$ with the $S_i$ half-shifts.}
We can then turn on ordinary discrete torsion for $\bZ_2^8$. The class controlling this discrete torsion is an element
\[
    \braket{-, -} \in H^2(B\bZ_2^8; U(1)) = \bZ_2^{28}
\]
specified by the $8(8-1)/2 = 28$ $\bZ_2$-valued pairings
\begin{equation} \label{eq:28pairs}
    \begin{aligned}
        \braket{S_i, S_j} ,&\qquad 1 \leq i < j \leq 6\,; \\
        \braket{\alpha, S_i}, \, \braket{\beta, S_i}, &\qquad 1 \leq i \leq 6\,; \\
        \braket{\alpha, \beta}.
    \end{aligned}
\end{equation}
Under the natural map, this then induces a generalised discrete torsion
\begin{equation} \label{eq:t6h}
    \omega \in H^2_{\bZ_2^\alpha \times \bZ_2^\beta}(T^6; U(1)) = U(1)^3 \times \bZ_2^{19}
\end{equation}
for the whole orbifold, where we have also given the calculation of the relevant cohomology group from \cref{app:hcalc}. There, we also show that the $\bZ_2^{19}$ part of $\omega$ is given by the $19$ pairings
\begin{equation} \label{eq:19pairs}
    \begin{aligned}
        \braket{S_i, S_j}, &\qquad 1 \leq i < j \leq 6 \text{ and } (i, j) \neq (1, 2), (3, 4), (5, 6)\, ; \\
        \braket{\alpha, S_{1,2}}, \braket{\beta, S_{3,4}}, \braket{\alpha\beta, S_{5,6}}; \\
        \braket{\alpha, \beta}.
    \end{aligned}
\end{equation}
These $19$ pairings can be independently chosen by varying $\braket{-, -}$. We learn that our simple construction based on $\bZ_2^8$ discrete torsion in fact accesses \emph{all} generalised discrete torsion phases in \eqref{eq:t6h}, at least up to deformation, which is all that matters.

The pullbacks of $\omega$ to the various fixed loci discussed in \cref{sec:t6geom} are also of interest, as they will control how each part of the singular locus is resolved in the orbifold CFT. At each $\alpha$-fixed $T^2$, there is an induced class
\begin{equation} \label{eq:t6t2h}
    \iota_{T^2_\alpha(a_3,a_4,a_5,a_6)}^*(\omega) \in H^2_{\bZ_2^\alpha \times \bZ_2^\beta}(T^2; U(1)) = U(1) \times \bZ_2^3\,.
\end{equation}
This class is described as follows. Choose any lift of $T^2_\alpha(a_3,a_4,a_5,a_6)$ from $T^6$ to $\widetilde{T}^6$. Then since the lifted $T^2$ is now localised at $(\tilde{x}_3, \tilde{x}_4, \tilde{x}_5, \tilde{x}_6) = \tfrac{1}{4}(a_3, a_4, a_5, a_6)$ mod $\tfrac{1}{2}$, it is no longer mapped into itself by $\alpha$ and $\beta$, but instead by the elements $\{\tilde{\alpha}, \tilde{\beta}, S_1, S_2\}$ where $\tilde{\alpha} = \alpha S_3^{a_3} S_4^{a_4} S_5^{a_5} S_6^{a_6}$ and $\tilde{\beta} = \beta S_5^{a_5} S_6^{a_6}$ which can be interpreted as the local versions of $\alpha$ and $\beta$ at that fixed $T^2$. Then the discrete part of the class \eqref{eq:t6t2h} is given by
\begin{equation} \label{eq:t6t2aw}
    \left[\iota_{T^2_\alpha(a_3,a_4,a_5,a_6)}^*(\omega)\right] = \left( \braket{\tilde{\alpha}, S_1}, \braket{\tilde{\alpha}, S_2}, \braket{\tilde{\alpha}, \tilde{\beta}} \right)\,.
\end{equation}
Similarly at the $\beta$-fixed $T^2$s we have
\begin{equation} \label{eq:t6t2bw}
    \left[\iota_{T^2_\beta(a_1,a_2,a_5,a_6)}^*(\omega)\right] = \left( \braket{\tilde{\beta}, S_3}, \braket{\tilde{\beta}, S_4}, \braket{\tilde{\alpha}, \tilde{\beta}} \right)
\end{equation}
where $\tilde{\alpha} = \alpha S_5^{a_5} S_6^{a_6}$ and $\tilde{\beta} = \beta S_1^{a_1} S_2^{a_2} S_5^{a_5} S_6^{a_6}$, while at the $\alpha\beta$-fixed $T^2$s we have
\begin{equation} \label{eq:t6t2abw}
    \left[\iota_{T^2_{\alpha\beta}(a_1,a_2,a_3,a_4)}^*(\omega)\right] = \left( \braket{\tilde{\alpha}\tilde{\beta}, S_5}, \braket{\tilde{\alpha}\tilde{\beta}, S_6}, \braket{\tilde{\alpha}, \tilde{\beta}} \right)
\end{equation}
where $\tilde{\alpha} = \alpha S_3^{a_3} S_4^{a_4}$ and $\tilde{\beta} = \beta S_1^{a_1} S_2^{a_2}$.

At each of the $64$ fixed points there is an induced local discrete torsion in
\begin{equation} \label{eq:t6pth}
    \iota_{({a_1,\dots,a_6)}/2}^*(\omega) \in H^2(B(\bZ_2^\alpha \times \bZ_2^\beta); U(1)) = \bZ_2
\end{equation}
whose value is given by $\braket{\tilde{\alpha}, \tilde{\beta}}$, where $\tilde{\alpha} = \alpha S_3^{a_3} S_4^{a_4} S_5^{a_5} S_6^{a_6}$ and $\tilde{\beta} = \beta S_1^{a_1} S_2^{a_2} S_5^{a_5} S_6^{a_6}$ are the local versions of $\alpha$ and $\beta$ at the fixed point.

Note that all induced discrete torsion phases $\iota_N^*(\omega)$ only depend on the $19$ pairings \eqref{eq:19pairs} that actually represent the class $\omega$.

\subsection{Local analysis} \label{sec:t6loc}

In this section we study the orbifold CFT $T^6 / \bZ_2^2$ with an arbitrary generalised discrete torsion, and compute the Hodge numbers $h^{1,1}$ and $h^{2,1}$.

We start by defining the flat sigma model on $\widetilde{T}^6$. This is $6$ copies of the $X\psi\bar{\psi}$ CFT, and consists of compact real scalars $X_i \sim X_i + 1$ and Majorana fermions $(\psi_i, \bar{\psi}_i)$ with the standard action
\[
    S_{(X\psi\bar{\psi})^6} = \int \! d^2x \left( \partial X_i \bar{\partial} X_i + \psi_i \bar{\partial} \psi_i + \bar{\psi}_i \partial \bar{\psi}_i \right)\,.
\]
We will consider the theory on a spatial circle with spin structure $\rho$. We then wish to gauge the $\bZ_2^8$ symmetry generated by $S_1, \dots, S_6, \alpha \coloneqq R_3 R_4 R_5 R_6, \beta \coloneqq R_1 R_2 R_5 R_6$ with ordinary discrete torsion $\braket{-, -}$, where $R_i$ and $S_i$ are the reflection $(X_i, \psi_i, \bar{\psi}_i) \mapsto (-X_i, -\psi_i, -\bar{\psi}_i)$ and the half-shift $X_i \mapsto X_i + \frac{1}{2}$ of the $i$-th direction, respectively. Introduce the gauge field $a = (a_1, \dots, a_6, a_\alpha, a_\beta) \in \bZ_2^8$. The Hilbert space of the gauged theory is
\[
    \cH^{(X\psi\bar{\psi})^6 / \bZ_2^8}_\rho = \bigoplus_{a \in \bZ_2^8} \cH^{(X\psi\bar{\psi})^6}_{\rho, a} \big|_{Q = \braket{-, a}}
\]
which takes the form of a sum over all twisted sectors, where in each twisted sector $\cH^{(X\psi\bar{\psi})^6}_{\rho,a}$, we pick out the subspace of charge $\braket{-, a}$ selected by the twist $a$ and the discrete torsion $\braket{-, -}$. This sum can be written out more explicitly as
\begin{equation} \label{eq:t6Htwisted2}
    \cH^{(X\psi\bar{\psi})^6 / \bZ_2^8}_\rho = \bigoplus_{a \in \bZ_2^8} \bigoplus_{\substack{r \in \bZ_2^6 \\ r_3+r_4+r_5+r_6 = \braket{\alpha, a} \\ r_1+r_2+r_5+r_6 = \braket{\beta, a}}} \bigotimes_{i=1}^6 \cH^{X\psi\bar{\psi}}_{\rho, S^{a_i} R^{a_\alpha \delta_{i,3456} + a_\beta \delta_{i,1256}}} \big|_{\begin{subarray}{l} Q_S = \braket{S_i, a} \\ Q_R = r_i \end{subarray}}\,,
\end{equation}
where $\delta_{i,3456}=1$ or $0$ depending on whether $i$ is one of $3,4,5,6$ or not, 
and similarly for $\delta_{i,1256}$.

Our primary interest is when the spin structure $\rho$ is taken to be periodic (Ramond), and we focus only on the ground states, as the multiplicities of these (graded by $U(1)_L \times U(1)_R$ fermion number) are the Hodge numbers. So, the immediate task is to understand, for a single copy of the $X\psi\bar{\psi}$ CFT, which charge sectors of which twisted sectors under $\bZ_2^R \times \bZ_2^S$ contain ground states.

\begin{table}
    \centering
    \renewcommand{\arraystretch}{1.2}
    \begin{tabular}{c|c|c|c|c|c|}
        \multicolumn{1}{c}{} & \multicolumn{1}{c}{} & \multicolumn{4}{c}{twist around spatial circle} \\
        \cline{3-6}
        \multicolumn{1}{c}{} & & $1$ & $S$ & $R$ & $SR$ \\
        \cline{2-6}
        \multirow{4}{*}{charges} & $Q_S=0,Q_R=0$ & $|0\rangle_{\psi\bar{\psi}}$ & & $\tfrac{1}{\sqrt{2}}(|0\rangle_X + |\frac{1}{2}\rangle_X)$ & $\tfrac{1}{\sqrt{2}}(|\tfrac{1}{4}\rangle_X + |\tfrac{3}{4}\rangle_X)$ \\
        \cline{2-6}
        & $Q_S=1,Q_R=0$ & & & $\tfrac{1}{\sqrt{2}}(|0\rangle_X - |\frac{1}{2}\rangle_X)$ & \\
        \cline{2-6}
        & $Q_S=0,Q_R=1$ & $|1\rangle_{\psi\bar{\psi}}$ & & & \\
        \cline{2-6}
        & $Q_S=1,Q_R=1$ & & & & $\tfrac{1}{\sqrt{2}}(|\tfrac{1}{4}\rangle_X - |\tfrac{3}{4}\rangle_X)$ \\
        \cline{2-6}
    \end{tabular}
    \caption{The vacua of a single copy of the $X\psi\bar{\psi}$ CFT in the Ramond sector.}
    \label{tab:vacua}
\end{table}

A summary of the answer is shown in \cref{tab:vacua}. First, consider the case where the twist around the spatial circle is $1$ or $S$. Then the fermions are periodic, hence there are two vacua $\ket{0}_{\psi\bar{\psi}}$ and $\ket{1}_{\psi\bar{\psi}}$ with $\Delta = \frac{1}{8}$ distinguished by their fermion parity $(-1)^F$. Meanwhile the scalar has a unique ground state with $\Delta = 0$ in the untwisted sector and only excited states in the $S$-twisted sector. This explains the first two columns of the table. Next, consider the case where the twist is $R$ or $SR$. Then the fermions are antiperiodic, so have a unique vacuum with $\Delta = 0$, while the scalar has vacua $\ket{f}_X$ labelled by the fixed points of $R$ and $SR$, namely $f \in \{0,\tfrac{1}{2}\}$ for $R$ and $f \in \{\tfrac{1}{4}, \tfrac{3}{4}\}$ for $SR$, with $\Delta = \tfrac{1}{8}$. These are acted on as $R\ket{f}_X = \ket{-f}_X$ and $S\ket{f}_X = \ket{f+\tfrac{1}{2}}_X$, which explains the last two columns of the table.

We can now return to \eqref{eq:t6Htwisted2} and determine the ground states. In view of \cref{tab:vacua}, these can only come from pairs $a = (a_1, \dots, a_6, a_\alpha, a_\beta) \in \bZ_2^8$ and $r = (r_1, \dots, r_6) \in \bZ_2^6$ obeying the selection rules
\begin{equation}
    a_\alpha \delta_{i,3456} + a_\beta \delta_{i,1256}
    =
    \begin{cases}
        0 & \implies a_i = \braket{S_i, a} = 0 \,,\\
        1 & \implies r_i = a_i \braket{S_i, a}\,.
    \end{cases}
\end{equation}
Solving these together with the constraints already placed on $r$ in \eqref{eq:t6Htwisted2}, the solutions fall into the following classes:
\begin{itemize}

\item Solutions with $(a_\alpha, a_\beta) = (0, 0)$. These are parameterised by tuples $(r_1, \dots, r_6)$ satisfying $r_3+r_4+r_5+r_6 = 0$ and $r_1+r_2+r_5+r_6 = 0$. Each solution contributes a ground state $\bigotimes_{i=1}^6 \ket{r_i}_{\psi_i \bar{\psi}_i}$. The states can be displayed in the Hodge diamond
\[
    \begin{array}{ccccccc}
        & & & 1 \\
        & & 0 & & 0 \\
        & 0 & & 3 & & 0 \\
        1 & & 3 & & 3 & & 1 \\
        & 0 & & 3 & & 0 \\
        & & 0 & & 0 \\
        & & & 1
    \end{array}
\]
upon identifying the $U(1)_L \times U(1)_R$ fermion number, suitably shifted by $(+3/2, +3/2)$, with the Dolbeault degree $(p, q)$. This part describes the classical orbifold $T^6 / \bZ_2^2$.

\item Solutions with $(a_\alpha, a_\beta) = (1, 0)$. These are parameterised by tuples $(r_1, r_2, a_3, a_4, a_5, a_6)$, where the part $(a_3, a_4, a_5, a_6)$ satisfies the constraint $\braket{\tilde{\alpha}, S_{1,2}} = 0$ where $\tilde{\alpha} = \alpha S_3^{a_3} S_4^{a_4} S_5^{a_5} S_6^{a_6}$, and the part $(r_1, r_2)$ satisfies $r_1 + r_2 = \braket{\tilde{\alpha}, \tilde{\beta}}$ where $\tilde{\beta} = \beta S_5^{a_5} S_6^{a_6}$. Each solution contributes a ground state with fermionic part $\bigotimes_{i=1,2} \ket{r_i}_{\psi_i \bar{\psi}_i}$. The states coming from each admissible tuple $(a_3, a_4, a_5, a_6)$ can be displayed in the Hodge diamond
\[
    \begin{array}{c}
        \begin{array}{ccccccc}
            & & & 0 \\
            & & 0 & & 0 \\
            & 0 & & 1 & & 0 \\
            0 & & 0 & & 0 & & 0 \\
            & 0 & & 1 & & 0 \\
            & & 0 & & 0 \\
            & & & 0
        \end{array} \vspace{3ex} \\
        \text{if } \braket{\tilde{\alpha}, \tilde{\beta}} = 0
    \end{array}
    \qquad
    \begin{array}{c}
        \begin{array}{ccccccc}
            & & & 0 \\
            & & 0 & & 0 \\
            & 0 & & 0 & & 0 \\
            0 & & 1 & & 1 & & 0 \\
            & 0 & & 0 & & 0 \\
            & & 0 & & 0 \\
            & & & 0
        \end{array} \vspace{3ex} \\
        \text{if } \braket{\tilde{\alpha}, \tilde{\beta}} = 1
    \end{array}
\]
Since each $(a_3, a_4, a_5, a_6)$ labels an $\alpha$-fixed $T^2$, we interpret this calculation as telling us that this $\alpha$-fixed $T^2$ remains singular if the part $\braket{\tilde{\alpha}, S_{1,2}}$ of the local discrete torsion \eqref{eq:t6t2aw} is non-zero, otherwise, it is resolved or deformed depending on whether the part $\braket{\tilde{\alpha}, \tilde{\beta}}$ takes the value $0$ or $1$ respectively.

\item The structure of the remaining solutions are identical to those above, so we will be brief.

\begin{itemize}

\item The solutions with $(a_\alpha, a_\beta) = (0, 1)$ are parameterised by $(a_1, a_2, r_3, r_4, a_5, a_6)$ satisfying $\braket{\tilde{\beta}, S_{3,4}} = 0$ and $r_3 + r_4 = \braket{\tilde{\alpha}, \tilde{\beta}}$ where $\tilde{\alpha} = \alpha S_5^{a_5} S_6^{a_6}$ and $\tilde{\beta} = \beta S_1^{a_1} S_2^{a_2} S_5^{a_5} S_6^{a_6}$, and represent the $\beta$-fixed tori.

\item The solutions with $(a_\alpha, a_\beta) = (1, 1)$ are parameterised by $(a_1, a_2, a_3, a_4, r_5, r_6)$ satisfying $\braket{\tilde{\alpha}\tilde{\beta}, S_{5,6}} = 0$ and $r_5 + r_6 = \braket{\tilde{\alpha}, \tilde{\beta}}$ where $\tilde{\alpha} = \alpha S_3^{a_3} S_4^{a_4}$ and $\tilde{\beta} = \beta S_1^{a_1} S_2^{a_2}$, and represent the $\alpha\beta$-fixed tori.

\end{itemize}

\end{itemize}
Summing the contributions from all ground states, we obtain the following Hodge numbers, which forms the main result of this section:

\begin{claim}
    The Hodge numbers of the orbifold CFT $T^6 / \bZ_2^2$ with generalised discrete torsion $\omega$ are
    \[
        (h^{1,1}, h^{2,1}) = (3, 3) + \sum_{\scriptsize
        \begin{alignedat}{1}
            \alpha \text{-fixed } T^2 \text{s with } &\braket{\tilde{\alpha}, S_{1,2}} = 0 \\
            \beta \text{-fixed } T^2 \text{s with } &\braket{\tilde{\beta}, S_{3,4}} = 0 \\
            \alpha\beta \text{-fixed } T^2 \text{s with } &\braket{\tilde{\alpha}\tilde{\beta}, S_{5,6}} = 0
        \end{alignedat}
        }
        \begin{cases}
            (1, 0) & \text{if}\ \braket{\tilde{\alpha}, \tilde{\beta}} = 0 \\
            (0, 1) & \text{if}\ \braket{\tilde{\alpha}, \tilde{\beta}} = 1
        \end{cases}
    \]
    where the class $\omega$ is specified by the pairings \eqref{eq:19pairs}, and the meaning of $\tilde{\alpha}$ and $\tilde{\beta}$ depends on the fixed locus under consideration and is given in \cref{eq:t6t2aw,eq:t6t2bw,eq:t6t2abw}.
\end{claim}

It is interesting to ask when the classical orbifold $T^6 / \bZ_2^2$ is fully desingularised, that is $h^{1,1} + h^{2,1} = 54$. For this to happen, we must have
\[
    \begin{aligned}
        \braket{\alpha, S_{1,2}} &= \braket{S_{3,4,5,6}, S_{1,2}} = 0 \,,\\
        \braket{\beta, S_{3,4}} &= \braket{S_{1,2,5,6}, S_{3,4}} = 0 \,,\\
        \braket{\alpha\beta, S_{5,6}} &= \braket{S_{1,2,3,4}, S_{5,6}} = 0\,.
    \end{aligned}
\]
Together these are $18$ constraints on $19$ phases, and they set all of them to zero except for $\braket{\alpha, \beta}$. So one is only allowed to turn on ordinary discrete torsion, resulting in the Calabi--Yau manifolds with $(h^{1,1}, h^{2,1}) = (51, 3), (3, 51)$ already found by \cite{Vafa:1994rv}, not the full set $(51-3\ell, 3+3\ell)$ with $0 \leq \ell \leq 16$ described by \cite{Gaberdiel:2004vx}.

The above result is expected. It is not possible to independently choose to deform or blow up each $T^2$ even in classical geometry, because when two $T^2$s intersect we must make the same choice for both of them in order to maintain smoothness at their intersection. But the collection of all $T^2$s is connected. So to obtain a smooth geometry, we must make the same choice for all $T^2$s, corresponding to ordinary discrete torsion.

This can also be understood in generalised discrete torsion terms. The $\alpha$-fixed $T^2$ specified by $(a_3, a_4, a_5, a_6)$ contains the four fixed points $\tfrac{1}{2}(a_1, \dots, a_6)$ for varying $a_{1,2} \in \{0,1\}$. The local discrete torsion \eqref{eq:t6pth} at such a point is $\braket{\alpha S_3^{a_3} S_4^{a_4} S_5^{a_5} S_6^{a_6}, \beta S_1^{a_1} S_2^{a_2} S_5^{a_5} S_6^{a_6}}$. The condition for these to all be the same is $\braket{\alpha S_3^{a_3} S_4^{a_4} S_5^{a_5} S_6^{a_6}, S_{1,2}} = 0$, which is equivalent to the vanishing of the part $\braket{\tilde{\alpha}, S_{1,2}}$ of the local discrete torsion \eqref{eq:t6t2aw} at $T^2$.

\section{$T^7/\bZ_2^3$} \label{sec:t7}

In this section we consider a $G_2$ orbifold $T^7/\bZ_2^3$, considered by \cite{JoyceBook,Gaberdiel:2004vx}.

\subsection{Geometry} \label{sec:t7geom}

Let $x_1, \dots, x_7$ denote the seven periodic coordinates of $T^7$, with $x_i \sim x_i + 1$.
We then let the generators $\alpha$, $\beta$, $\gamma$ of $\bZ_2^3$ act by
\begin{equation} \label{eq:t7g}
    \begin{aligned}
        \alpha &\coloneqq R_1 R_2 R_3 R_4, &
        \beta &\coloneqq R_1 (SR)_2 R_5 R_6,&
        \gamma &\coloneqq (SR)_1 R_3 R_5 R_7,
    \end{aligned}
\end{equation}
where $R_i$ and $S_i$ are the reflection $x_i \mapsto -x_i$ and the half-shift $x_i \mapsto x_i + \frac12$  of the $i$-th direction, respectively.

The $\alpha$-fixed locus consists of $16$ copies of $T^3$ localised at $(x_1, x_2, x_3, x_4) = \tfrac{1}{2}(a_1, a_2, a_3, a_4)$ with $a_{1,2,3,4} \in \{0,1\}$. These $T^3$s are acted on freely by $\bZ_2^\beta \times \bZ_2^\gamma$, forming $4$ orbits of $4$ tori, where the orbits are labelled by $(a_3, a_4)$ and the tori within each orbit are labelled by $(a_1, a_2)$.

The $\beta$-fixed locus has an identical structure. It consists of $16$ copies of $T^3$ localised at $(x_1, x_2, x_5, x_6) = \tfrac{1}{2}(a_1, a_2 + \tfrac{1}{2}, a_5, a_6)$ with $a_{1,2,5,6} \in \{0,1\}$. These $T^3$s are acted on freely by $\bZ_2^\alpha \times \bZ_2^\gamma$, forming $4$ orbits of $4$ tori, where the orbits are labelled by $(a_5, a_6)$ and the tori within each orbit are labelled by $(a_1, a_2)$.

The $\gamma$-fixed locus consists of $16$ copies of $T^3$ localised at $(x_1, x_3, x_5, x_7) = \frac{1}{2} (a_1 + \frac{1}{2}, a_3, a_5, a_7)$ with $a_{1,3,5,7} \in \{0,1\}$. Each $T^3$ is mapped into itself by $\alpha\beta$, but otherwise acted on freely by $(\bZ_2^\alpha \times \bZ_2^\beta) / \bZ_2^{\alpha\beta} \cong \bZ_2$, forming $8$ orbits of $2$ tori, where the orbits are labelled by $(a_3, a_5, a_7)$ and the tori within each orbit are labelled by $a_1$.

There are no fixed points under the remaining elements $\alpha\beta$, $\alpha\gamma$, $\beta\gamma$.
This in particular means that the $\alpha$-, $\beta$- and $\gamma$-fixed loci do not intersect among themselves in this case.
This is in contrast to the situation in \cref{sec:t6}.

\subsection{Cohomology} \label{sec:t7coh}

To access a subset of the generalised discrete torsion phases for $T^7/\bZ_2^3$, we will write $T^7 = \widetilde{T}^7 / \bZ_2^5$ by doubling the radii of circles $3$-$7$ and gauging the $5$ half-shifts $S_3, \dots, S_7$. The orbifold $T^7/\bZ_2^3$ can then be rewritten as $\widetilde{T}^7/\bZ_2^8$ where the $\bZ_2^8$ symmetry is generated by $\{S_3,\dots,S_7,\alpha,\beta,\gamma\}$.%
\footnote{\label{foot:T7hatG}%
According to the general discussion in \cref{sec:gauging}, this is the unique rewriting of the orbifold involving the largest possible abelian group. Indeed, the group $\hat{G}$ such that $T^7/\bZ_2^3 = \bR^7/\hat{G}$ takes the form of an extension $1 \to \bZ^7 \to \hat{G} \to \bZ_2^3 \to 1$ that is not a semidirect product. To describe $\hat{G}$, let $T_{1,\dots,7}$ denote generators for the $\bZ^7$, and $\alpha,\beta,\gamma$ lifts of the generators of $\bZ_2^3$.
Then the structure of $\hat{G}$ is specified by the relations
\begin{align*}
    \alpha^2 = \beta^2 = \gamma^2 &= 1, &
    \alpha T_i \alpha^{-1} &= T_i^{(-1,-1,-1,-1,+1,+1,+1)_i}, \\
    \alpha\beta\alpha\beta &= T_2^{-1} ,&
    \beta T_i \beta^{-1} &= T_i^{(-1,-1,+1,+1,-1,-1,+1)_i},\\
    \alpha\gamma\alpha\gamma = \beta\gamma\beta\gamma &= T_1^{-1}, &
    \gamma T_i \gamma^{-1} &= T_i^{(-1,+1,-1,+1,-1,+1,-1)_i}.
\end{align*}
One then computes that the derived subgroup $\hat{G}'$ is generated by $T_1, T_2, T_3^2, \dots, T_7^2$, and therefore that $\hat{G}^\ab = \bZ_2^8$ generated by $T_3, \dots, T_7, \alpha, \beta, \gamma$. On $\widetilde{T}^7 = \mathbb{R}^7/\hat{G}'$, these generators act as $S_3, \dots, S_7, \alpha, \beta, \gamma$ with the $S_i$ half-shifts.}
We can then turn on ordinary discrete torsion for $\bZ_2^8$. As before, the class controlling this discrete torsion phase is an element
\[
    \braket{-, -} \in H^2(B\bZ_2^8; U(1)) = \bZ_2^{28}
\]
specified by the $8(8-1)/2 = 28$ $\bZ_2$-valued pairings
\begin{equation} \label{eq:28pairs2}
    \begin{aligned}
        \braket{S_i, S_j}, &\qquad 3 \leq i < j \leq 7\,; \\
        \braket{\alpha, S_i}, \, \braket{\beta, S_i}, \, \braket{\gamma, S_i}, &\qquad 3 \leq i \leq 7 \,;\\
        \braket{\alpha, \beta}, \, \braket{\alpha, \gamma}, \, \braket{\beta, \gamma}\,.
    \end{aligned}
\end{equation}
Under the natural map, this then induces a generalised discrete torsion
\begin{equation} \label{eq:t7h}
    \omega \in H_{\bZ_2^3}^2(T^7; U(1)) = \bZ_2^{21}
\end{equation}
for the whole orbifold, where we have also given the calculation of the relevant cohomology group from \cref{app:hcalc}. There, we also show that $\omega$ is given by the $21$ pairings
\begin{equation} \label{eq:21pairs}
    \begin{aligned}
        \braket{S_i, S_j} , &\qquad 3 \leq i < j \leq 7\,; \\
        \braket{\alpha, S_{5,6,7}}, \, \braket{\beta, S_{3,4,7}}, \, \braket{\gamma, S_{4,6}}, \, \braket{\alpha\gamma, S_3}, \, \braket{\beta\gamma, S_5} ; \\
        \braket{\alpha\beta, \gamma}.
    \end{aligned}
\end{equation}
These $21$ pairings can be independently chosen by varying $\braket{-, -}$. We learn that our simple construction based on $\bZ_2^8$ discrete torsion again accesses all generalised discrete torsion phases in \eqref{eq:t7h}.

Let us also consider the pullbacks of $\omega$ to the various fixed loci discussed in \cref{sec:t7geom}, as these will control how each part of the singular locus is resolved in the orbifold CFT. At each $\alpha$-fixed $T^3$, there is an induced class
\begin{equation} \label{eq:t7t3ah}
    \iota_{T^3_\alpha(a_1,a_2,a_3,a_4)}^*(\omega) \in H_{\bZ_2^\alpha}^2(T^3; U(1)) = U(1)^3 \times \bZ_2^3\,.
\end{equation}
This class is described as follows. Choose any lift of $T^3_\alpha(a_1,a_2,a_3,a_4)$ from $T^7$ to $\widetilde{T}^7$. Then since the lifted $T^3$ is now localised at $(\tilde{x}_1, \tilde{x}_2, \tilde{x}_3, \tilde{x}_4) = (\tfrac{a_1}{2}, \tfrac{a_2}{2}, \tfrac{a_3}{4} \text{ mod } \tfrac{1}{2}, \tfrac{a_4}{4} \text{ mod } \tfrac{1}{2})$, it is no longer mapped into itself by $\alpha$, but instead by the elements $\{\tilde{\alpha}, S_5, S_6, S_7\}$ where $\tilde{\alpha} = \alpha S_3^{a_3} S_4^{a_4}$ which can be interpreted as the local version of $\alpha$ at that fixed $T^3$. Then the discrete part of the class \eqref{eq:t7t3ah} is given by
\begin{equation} \label{eq:t7t3aw}
    \left[\iota_{T^3_\alpha(a_1,a_2,a_3,a_4)}^*(\omega)\right] = \braket{\tilde{\alpha}, S_{5,6,7}}\,.
\end{equation}
Similarly at the $\beta$-fixed $T^3$s we have
\begin{equation} \label{eq:t7t3bw}
    \left[\iota_{T^3_\beta(a_1,a_2,a_5,a_6)}^*(\omega)\right] = \braket{\tilde{\beta}, S_{3,4,7}}
\end{equation}
where $\tilde{\beta} = \beta S_5^{a_5} S_6^{a_6}$. Meanwhile, at each $\gamma$-fixed $T^3$, there is an induced class
\begin{equation} \label{eq:t7t3yh}
    \iota_{T^3_\gamma(a_1,a_3,a_5,a_7)}^*(\omega) \in H_{\bZ_2^{\alpha\beta} \times \bZ_2^\gamma}^2(T^3; U(1)) = U(1) \times \bZ_2^3\,.
\end{equation}
To describe this class, choose any lift of $T^3_\gamma(a_1,a_3,a_5,a_7)$ from $T^7$ to $\widetilde{T}^7$. Then since the lifted $T^3$ is now localised at $(\tilde{x}_1,\tilde{x}_3,\tilde{x}_5,\tilde{x}_7) = (\tfrac{a_1}{2}, \tfrac{a_3}{4} \text{ mod } \tfrac{1}{2}, \tfrac{a_5}{4} \text{ mod } \tfrac{1}{2}, \tfrac{a_7}{4} \text{ mod } \tfrac{1}{2})$, it is no longer mapped into itself by $\alpha\beta$ and $\gamma$, but instead by the elements $\{\widetilde{\alpha\beta}, \tilde{\gamma}, S_4, S_6\}$ where $\widetilde{\alpha\beta} = \alpha\beta S_3^{a_3} S_5^{a_5}$ and $\tilde{\gamma} = \gamma S_3^{a_3} S_5^{a_5} S_7^{a_7}$. Then the discrete part of the class \eqref{eq:t7t3yh} is given by
\begin{equation} \label{eq:t7t3yw}
    \left[\iota_{T^3_\gamma(a_1,a_3,a_5,a_7)}^*(\omega)\right] = \left( \braket{\tilde{\gamma}, S_4}, \braket{\tilde{\gamma}, S_6}, \braket{\widetilde{\alpha\beta}, \tilde{\gamma}} \right)\,.
\end{equation}
Note that all induced discrete torsion phases $\iota_N^*(\omega)$ only depend on the $21$ pairings \eqref{eq:21pairs} that actually represent the class $\omega$.

\subsection{Local analysis} \label{sec:t7loc}

In this section we study the orbifold CFT $T^7 / \bZ_2^3$ with an arbitrary generalised discrete torsion, and compute the Betti numbers $b_2, b_3$.

For this we largely follow the same story reviewed in \cref{sec:t6loc}. We take the flat sigma model on $\widetilde{T}^7$, which consists of $7$ copies of the $X\psi\bar{\psi}$ CFT, place it on a spatial circle with the spin structure $\rho$, and gauge the $\bZ_2^8$ symmetry generated by $S_3, \dots, S_7, \alpha \coloneqq R_1 R_2 R_3 R_4, \beta \coloneqq R_1 (SR)_2 R_5 R_6, \gamma \coloneqq (SR)_1 R_3 R_5 R_7$ with ordinary discrete torsion $\braket{-, -}$. Introduce the gauge field $a = (a_3, \dots, a_7, a_\alpha, a_\beta, a_\gamma) \in \bZ_2^8$. Then the Hilbert space of the gauged theory is
\begin{equation} \label{eq:t7Htwisted2}
    \cH^{(X\psi\bar{\psi})^7 / \bZ_2^8}_\rho = \bigoplus_{a \in \bZ_2^8} \bigoplus_{\substack{r \in \bZ_2^7 \\ r_1+r_2+r_3+r_4 = \braket{\alpha, a}}}
    \left[
    \begin{aligned}
        &\cH^{(X\psi\bar{\psi})_1}_{\rho, S^{a_\gamma}R^{a_\alpha+a_\beta+a_\gamma}}
        \big|_{\begin{subarray}{l} Q_S = r_1+r_3+r_5+r_7 + \braket{\gamma, a} \\ Q_R = r_1 \end{subarray}}
        \\
        \otimes &\cH^{(X\psi\bar{\psi})_2}_{\rho, S^{a_\beta}R^{a_\alpha+a_\beta}}
        \big|_{\begin{subarray}{l} Q_S = r_1+r_2+r_5+r_6 + \braket{\beta, a} \\ Q_R = r_2 \end{subarray}}
        \\
        \otimes &\bigotimes_{i=3}^7 \cH^{(X\psi\bar{\psi})_i}_{\rho, S^{a_i} R^{a_\alpha \delta_{i,1234} + a_\beta \delta_{i,1256} + a_\gamma \delta_{i,1357}}} \big|_{\begin{subarray}{l} Q_S = \braket{S_i, a} \\ Q_R = r_i \end{subarray}}
    \end{aligned}
    \right]\,.
\end{equation}
We are solely interested in the vacua in the Ramond sector $\rho = \rho_R$. Using \cref{tab:vacua}, the pairs $a = (a_3, \dots, a_7, a_\alpha, a_\beta, a_\gamma) \in \bZ_2^8$ and $r = (r_1, \dots, r_7) \in \bZ_2^7$ that contribute vacua are those obeying the selection rules
\begin{align}
    a_\alpha+a_\beta+a_\gamma
    &=
    \begin{cases}
        0 & \implies a_\gamma = \sum_{i=1,3,5,7} r_i + \braket{\gamma, a} = 0 \,,\\
        1 & \implies r_1 = a_\gamma (\sum_{i=1,3,5,7} r_i + \braket{\gamma, a})\,;
    \end{cases}
    \\
    a_\alpha+a_\beta
    &=
    \begin{cases}
        0 & \implies a_\beta = \sum_{i=1,2,5,6} r_i + \braket{\beta, a} = 0 \,,\\
        1 & \implies r_2 = a_\beta (\sum_{i=1,2,5,6} r_i + \braket{\beta, a})\,;
    \end{cases}
    \\
    a_\alpha \delta_{i,1234} + a_\beta \delta_{i,1256} + a_\gamma \delta_{i,1357}
    &=
    \begin{cases}
        0 & \implies a_i = \braket{S_i, a} = 0 \,, \\
        1 & \implies r_i = a_i \braket{S_i, a}\,.
    \end{cases}
    \qquad
    \{i = 3,\dots,7\}
\end{align}
Solving these together with the constraint already placed on $r$ in \eqref{eq:t7Htwisted2}, the solutions fall into the following classes:
\begin{itemize}

\item Solutions with $(a_\alpha, a_\beta, a_\gamma) = (0, 0, 0)$. These are parameterised by tuples $(r_1, \dots, r_7)$ satisfying $\sum_{i=1,2,3,4} r_i = 0$, $\sum_{i=1,2,5,6} r_i = 0$, and $\sum_{i=1,3,5,7} r_i = 0$. Each solution contributes a ground state $\bigotimes_{i=1}^7 \ket{r_i}_{\psi_i \bar{\psi}_i}$. These states can be arranged into the spectrum of Betti numbers
\[
    (b_0, \dots, b_7) = (1, 0, 0, 7, 7, 0, 0, 1)
\]
by identifying $\sum_{i=1}^7 r_i$ with the de Rham degree. This part describes the classical orbifold $T^7 / \bZ_2^3$.

\item Solutions with $(a_\alpha, a_\beta, a_\gamma) = (1, 0, 0)$. These are parameterised by tuples $(a_3, a_4, r_5, r_6, r_7)$, where the part $(a_3, a_4)$ satisfies the constraint $\braket{\tilde{\alpha}, S_{5,6,7}} = 0$ where $\tilde{\alpha} = \alpha S_3^{a_3} S_4^{a_4}$. Each solution contributes a ground state with fermionic part $\bigotimes_{i=5,6,7} \ket{r_i}_{\psi_i \bar{\psi}_i}$. The states coming from each admissible tuple $(a_3, a_4)$ can be arranged into the spectrum of Betti numbers
\[
    (b_0, \dots, b_7) = (0, 0, 1, 3, 3, 1, 0, 0).
\]
Since $(a_3, a_4)$ labels a single orbit of $\alpha$-fixed $T^3$s, which corresponds to one singular locus in the classical orbifold, we interpret this calculation as telling us that this singular locus stays singular if the local discrete torsion \eqref{eq:t7t3aw} is non-zero, otherwise it is resolved in the unique way given above.

\item The structure of the solutions with $(a_\alpha, a_\beta, a_\gamma) = (0, 1, 0)$ is identical to those above, so we will be brief. The solutions are parameterised by $(r_3, r_4, a_5, a_6, r_7)$ satisfying $\braket{\tilde{\beta}, S_{3,4,7}} = 0$ where $\tilde{\beta} = \beta S_5^{a_5} S_6^{a_6}$, and represent the orbits of $\beta$-fixed tori.

\item Solutions with $(a_\alpha, a_\beta, a_\gamma) = (0, 0, 1)$. These are parameterised by tuples $(r_2, a_3, r_4, a_5, r_6, a_7)$, where the part $(a_3, a_5, a_7)$ satisfies the constraint $\braket{\tilde{\gamma}, S_{4,6}} = 0$ where $\tilde{\gamma} = \gamma S_3^{a_3} S_5^{a_5} S_7^{a_7}$, and the part $(r_4, r_6)$ satisfies $r_4 + r_6 = \braket{\widetilde{\alpha\beta}, \tilde{\gamma}}$ where $\widetilde{\alpha\beta} = \alpha\beta S_3^{a_3} S_5^{a_5}$. Each solution contributes a ground state with fermionic part $\bigotimes_{i=2,4,6} \ket{r_i}_{\psi_i\bar{\psi}_i}$. The states coming from each admissible tuple $(a_3, a_5, a_7)$ can be arranged into the spectrum of Betti numbers
\[
    (b_0, \dots, b_7) =
    \begin{cases}
        (0, 0, 1, 1, 1, 1, 0, 0)\,, & \braket{\widetilde{\alpha\beta}, \tilde{\gamma}} = 0 \,;\\
        (0, 0, 0, 2, 2, 0, 0, 0)\,, & \braket{\widetilde{\alpha\beta}, \tilde{\gamma}} = 1\,.
    \end{cases}
\]
Since $(a_3, a_5, a_7)$ labels a single orbit of $\gamma$-fixed $T^3$s, which corresponds to one singular locus in the classical orbifold, we interpret this calculation as telling us that this singular locus remains singular if the part $\braket{\tilde{\gamma}, S_{4,6}}$ of the local discrete torsion \eqref{eq:t7t3yw} is non-zero, otherwise, it is resolved in the two different ways above depending on the value of $\braket{\widetilde{\alpha\beta}, \tilde{\gamma}}$.

\end{itemize}
Summing the contributions from all ground states, we obtain the following Betti numbers, which forms the main result of this section:

\begin{claim}
    The Betti numbers of the orbifold CFT $T^7 / \bZ_2^3$ with generalised discrete torsion $\omega$ are
    \begin{align*}
        (b_2, b_3) = (0, 7)
        &+ \sum_{\scriptsize
        \begin{alignedat}{1}
            \alpha \text{-fixed orbits with } &\braket{\tilde{\alpha}, S_{5,6,7}} = 0 \\
            \beta \text{-fixed orbits with } &\braket{\tilde{\beta}, S_{3,4,7}} = 0
        \end{alignedat}
        } (1, 3) \\
        &+ \sum_{\scriptsize
        \begin{alignedat}{1}
            \gamma \text{-fixed orbits with } &\braket{\tilde{\gamma}, S_{4,6}} = 0
        \end{alignedat}
        }
        \begin{cases}
            (1, 1) \,,& \braket{\widetilde{\alpha\beta}, \tilde{\gamma}} = 0 \,,\\
            (0, 2) \,,& \braket{\widetilde{\alpha\beta}, \tilde{\gamma}} = 1 \,,
        \end{cases}
    \end{align*}
    where the class $\omega$ is specified by the pairings \eqref{eq:21pairs}, and the meaning of the tilded generators depends on the fixed locus under consideration and is given in \cref{eq:t7t3aw,eq:t7t3bw,eq:t7t3yw}.
\end{claim}

It is again interesting to ask when the geometry is fully desingularised, that is $b_2 + b_3 = 55$. For this to happen, we must have
\begin{equation} \label{eq:t7smoothness}
    \begin{aligned}
        \braket{\alpha, S_{5,6,7}} &= \braket{S_{3,4}, S_{5,6,7}} = 0 \,,\\
        \braket{\beta,  S_{3,4,7}} &= \braket{S_{5,6}, S_{3,4,7}} = 0 \,,\\
        \braket{\gamma, S_{4,6}}   &= \braket{S_{3,5,7}, S_{4,6}} = 0 \,.
    \end{aligned}
\end{equation}
Together these are $18$ constraints on $21$ phases, and they set all of them to zero except $\braket{\alpha\gamma, S_3}$, $\braket{\beta\gamma, S_5}$, and $\braket{\alpha\beta, \gamma}$. Thus things are more interesting than in the $T^6/\bZ_2^2$ case, because we now have the possibility of turning on generalised discrete torsion that is not simply ordinary discrete torsion.

To see how these phases affect the topology, we compute that at an orbit of $\gamma$-fixed $T^3$s,
\begin{equation} \label{eq:aby}
    \braket{\widetilde{\alpha\beta}, \tilde{\gamma}} = \braket{\alpha\beta, \gamma} + a_3 \braket{\alpha\gamma, S_3} + a_5 \braket{\beta\gamma, S_5}\,.
\end{equation}
Thus as $(a_3, a_5, a_7)$ takes all $8$ possible values, this expression is either all $0$s, all $1$s, or exactly half of both. Hence the total Betti numbers are
\begin{align}
    (b_2, b_3) = (8 + \ell, 47 - \ell)
    \quad \text{with} \quad
    \ell =
    \begin{cases}
        4\,, & \braket{\alpha\gamma, S_3} \neq 0 \text{ or } \braket{\beta\gamma, S_5} \neq 0 \,,\\
        8 \,,& \braket{\alpha\beta, \gamma} = 0 \,,\\
        0 \,,& \braket{\alpha\beta, \gamma} = 1\,.
    \end{cases}
\end{align}
As in the $T^6/\bZ_2^2$ case, this is only a subset of the Betti numbers $0 \leq \ell \leq 8$ described by \cite{Gaberdiel:2004vx}, but this time it is for a different reason. For $T^6/\bZ_2^2$, we were forced to either blow up or resolve all singular $T^2$s simultaneously, because of consistency at their intersections. But here there are no intersections among the singular $T^3$s. Instead the issue is that the local discrete torsions \eqref{eq:aby} cannot be chosen independently. Note that even if we were to relax the constraint \eqref{eq:t7smoothness} and allow arbitrary $\omega$, the local discrete torsions $\braket{\widetilde{\alpha\beta}, \tilde{\gamma}}$ still cannot be chosen independently: only an even number of them can be $1$.

This poses an interesting puzzle for us, since it is known at the level of classical geometry that one \emph{can} independently resolve each $\gamma$-fixed singular $T^3/\bZ_2$ in the orbifold. Indeed, this is exactly how mathematicians construct compact $G_2$ manifolds corresponding to all $0 \leq \ell \leq 8$. \cite{Joyce:1998en} What our investigation tells us is that the worldsheet description of these $G_2$ manifolds is not captured by an orbifold CFT with generalised discrete torsion taking values in $H_{\bZ_2^3}^2(T^7; U(1))$.

\section*{Acknowledgments}

YT would like to thank his late supervisor, Tohru Eguchi, for mentioning his desire to understand
the higher-genus consistency of the generalised discrete torsion of \cite{Gaberdiel:2004vx},
after YT gave a journal club on that paper as a graduate student on October 22nd, 2004.
His comment lingered in YT's ears for more than two decades,
which eventually led to the result presented in this paper.
It is a pity that YT does not have any way to communicate this development to him.

The authors would also like to thank M.~Gaberdiel and E.~Sharpe for providing valuable feedback on the draft of the paper.

YT is supported in part by WPI Initiative, MEXT, Japan at Kavli IPMU, the University of Tokyo
and by JSPS KAKENHI Grant-in-Aid (Kiban-C), No.24K06883.
PBS is supported by the ERC-COG grant NP-QFT No.~864583, by the MUR-FARE2020 grant No.~R20E8NR3HX, and for part of this work also by WPI Initiative, MEXT, Japan at Kavli IPMU, the University of Tokyo.

\appendix
\crefalias{section}{appendix}

\section{Comparison to the approach by Sharpe} \label{app:sharpe}

Here, we give evidence that the generalised discrete torsion
specified by the equivariant cohomology group $H^2_G(M;U(1))$ which was discussed in this paper
reproduces the flat part of the equivariant gerbe construction of Sharpe,
by considering the case of finite-group orbifolds of $M=T^n$ concretely.

In the bottom of page 7 of \cite{Sharpe:2003cs}, it is stated that the phase
introduced from an equivariant gerbe is given by \begin{equation}
\exp\left(
\int_0^{2\pi} \Lambda(g)_i \frac{\partial X^i}{\partial \sigma} d\sigma-
\int_0^{2\pi} \Lambda(h)_i \frac{\partial X^i}{\partial \tau} d\tau
\right)
\label{sharpe}
\end{equation}
where $X$ is the sigma model field,
and $\Lambda(g)_i$ are certain $U(1)$ holonomies along the direction $dX^i$ depending on $g\in G $,
which is a part of the data of a $G$-equivariant gerbe on $T^n$.
This phase \eqref{sharpe} is to be added to the standard contribution $\langle g,h\rangle=\omega(g,h)- \omega(h,g)$,
where $\omega$ is the 2-cocycle.

To see that we have the same phase from our approach,
we interpret $H^2(BG ,U(1))$ as specifying a $U(1)$-extension of $G$.
Before talking about the case of $\bZ^n \rtimes G$, consider the case of $G$.
Consider an extension
\begin{equation}
1\to U(1)\to \tilde{G} \to G \to 1.
\end{equation}
For $g\in G $, denote a lift in $\tilde{G} $ by $X_g$.
The 2-cocycle appears as
\begin{equation}
X_{g} X_h = e^{2\pi \ii \omega(g,h)} X_{gh}.
\end{equation}
The twists $g,h$ in the $T^2$ partition function commute.
In this case, \begin{equation}
X_h X_g= e^{2\pi \ii \omega(h,g)} X_{hg}= e^{2\pi \ii \omega(h,g)}X_{gh},
\end{equation}
and therefore
\begin{equation}
X_g X_h = e^{2\pi \ii \langle g,h\rangle} X_h X_g.
\end{equation}
From this reason $\langle g,h\rangle$ is also known as the commutator function of this central extension.

We now consider the $U(1)$ extension $\tilde \Gamma$ of $\Gamma\coloneqq\bZ^n\rtimes G $.
To fix the notations, denote the generators of $\bZ^n$ by $e_{1,\ldots,n}$,
and a generic element of $\bZ^n$ by $a=\sum_i a^i e_i$ with $a^i\in \bZ$.
We denote the right action of $g\in G $ by $a\mapsto ag$.
We denote the elements of $\Gamma$ as $y_a g$ for $a\in \bZ^d$ and $g\in G $,
so that $y_ay_b=y_{a+b}$, $y_a g= g y_{ag}$.
The symbol $y$ is here to convert an additive group $\bZ^d$ into a multiplicative notation.

Now, to study the structure of $\tilde \Gamma$, we still define $X_g$ as before,
satisfying $X_g X_h = \omega(g,h)X_{gh}$.
Denote the lifts of $e_i$ by $T_{1,\ldots,n}$.
Ordinary $B$-fields appear as phases $T_s T_t = e^{2\pi \ii B_{st}} T_s T_t$.
Let us now fix the lifts of $a=\sum_i a^i e_i\in \bZ^n$ to be \begin{equation}
Y_{a} \coloneqq (T_1)^{a^1} (T_2)^{a^2} \cdots (T_n)^{a^n}.
\end{equation}
Recall that every element in $\bZ^n\rtimes G $ is of the form $y_ag$ for $a\in \bZ^n$ and $g\in G $.
Fix their lifts to be $Y_a X_g$.

Then, to fix the structure of $\tilde \Gamma$, we need to specify how we can rearrange
 $Y_a X_g Y_b X_h$ into $Y_{a+bg} X_g X_h$.
For this we only have to fix the phases appearing in \begin{equation}
Y_{e_i} X_g = e^{2\pi \ii \Lambda(g)_i } X_g Y_{(e_i)g}.
\end{equation}
So the total data necessary to specify the extension are: $\omega(g,h)$, $B_{st}$, and $\Lambda(g)_i$,
satisfying some more constraints.

Now, consider the $T^2$ partition function with twists $y_ag$ and $y_bh$.
In the notation of Sharpe,
\begin{equation}
a^i = \int \frac{\partial X^i}{\partial \tau} d\tau,\qquad
b^i = \int \frac{\partial X^i}{\partial \sigma} d\sigma.
\end{equation}
Then the associated phase is the commutator function appearing in the comparison between
$(Y_a X_g) ( Y_b X_h)$ and $( Y_b X_h) (Y_a X_g) $.

On $T^2$, $y_ag$ and $y_bh$ need to commute.
In addition, for the consistency of the map $X: T^2\to T^n/G $,
$b$ needs to be fixed by $g$ and $a$ needs to be fixed by $h$.
Assume for simplicity that $B_{st}=0$.
Then we have \begin{align}
(Y_a X_g) ( Y_b X_h) &= e^{2\pi \ii \omega(g,h)}e^{2\pi\ii\Lambda(g)_i b^i} Y_{a+b} X_{gh},\\
(Y_b X_h) ( Y_a X_g) &= e^{2\pi \ii \omega(h,g)}e^{2\pi\ii\Lambda(h)_i a^i} Y_{a+b} X_{hg}.
\end{align}
Therefore the commutator phase is \begin{equation}
\omega(g,h)-\omega(h,g)+ (\Lambda(g)_i b^i-\Lambda(h)_i a^i),
\end{equation} exactly the form discussed by Sharpe.

\section{Cohomology calculations} \label{app:hcalc}

We compute the various equivariant cohomology groups used in \cref{sec:t6,sec:t7} in this appendix.
We employ two independent methods to check the computations.
First, we will simply construct explicit cell complexes for all spaces involved.
Second, we will use the correspondence to central extension of groups.

\subsection{Via cell complexes}
First, let us recall the standard computation of the group cohomology of $\bZ_2$. The space $E\bZ_2$ has two $n$-cells $\{a_n, t a_n\}$ in each dimension, where $t$ denotes the $\bZ_2$ generator. The differential is
\[
    \partial a_n = (1 + (-1)^n t) a_{n-1}
\]
and commutes with $t$.\footnote{This structure arises by considering $E\bZ_2 = S^\infty$ with $t$ acting by antipodal reflection, whereupon the two $n$-cells are the two hemispheres of $S^n$.} We would like to form $B\bZ_2 = E\bZ_2 / \bZ_2$. To do this, we pick one cell from each $\bZ_2$ orbit, namely $a_n$, and declare these to be the cells of $B\bZ_2$. To define the differential, we first act with the original differential, which produces some cells not in $B\bZ_2$, then use the $\bZ_2$ action to bring those cells back into $B\bZ_2$. This defines the new cell complex
\[
    \partial a_n = (1 + (-1)^n) a_{n-1}
\]
which computes the group cohomology of $\bZ_2$. For more details, see \cite{brown2012coh}.

The Borel quotients relevant to our torus orbifolds can be dealt with in a similar way, using the above calculation as a building block. For $(T^6 \times E\bZ_2^2) / \bZ_2^2$, we take $2$ copies of $E\bZ_2$ and $6$ copies of $S^1$ with the cell structure
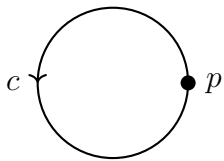
\[
    \begin{tikzpicture}
        \fill[black] (1,0) circle [radius=0.1];
        \draw[thick,decoration={markings,mark=at position 0.5 with {\arrow{>}}},postaction={decorate}] (1,0) arc[start angle=0, delta angle=360, radius=1];
        \node at (1,0) [anchor=west,outer sep=0.1cm] {$p$};
        \node at (-1,0) [anchor=east,outer sep=0.1cm] {$c$};
    \end{tikzpicture}
\]
where it is understood that the reflection $R$ acts as $c \mapsto -c$, $p \mapsto p$. Then performing the quotient by $\bZ_2^2$ yields a cell structure with cells
\[
    a_n b_m c^k \; : \; n,m \geq 0, k \in \{0,1\}^6
\]
and differentials
\begin{align*}
    \partial a_n &= (1+(-1)^n \alpha) a_{n-1} \,,\\
    \partial b_m &= (1+(-1)^m \beta) b_{m-1} \,,\\
    \partial c^k &= 0
\end{align*}
where $\alpha = (-1)^{k_3+k_4+k_5+k_6}$ and $\beta = (-1)^{k_1+k_2+k_5+k_6}$. Here $c^k = c_1^{k_1} \dots c_6^{k_6}$ denotes the general cell of $T^6$, where we suppress zero-cells in our notation.

For $(T^7 \times E\bZ_2^3) / \bZ_2^3$, note that the $\bZ_2^2$ subgroup $\{1,\alpha\beta,\alpha\gamma,\beta\gamma\}$ acts freely, and so the Borel quotient is homotopy equivalent to the simpler space $(T^7 \times  E\bZ_2) / \bZ_2^3$ where all $3$ $\bZ_2$ factors act on $E\bZ_2$. For the first $2$ circles of $T^7$, we take the cell structure
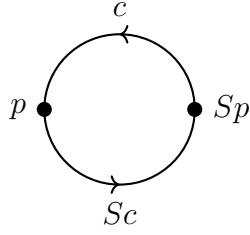
\[
    \begin{tikzpicture}
        \fill[black] (1,0) circle [radius=0.1];
        \fill[black] (-1,0) circle [radius=0.1];
        \draw[thick,decoration={markings,mark=at position 0.5 with {\arrow{>}}},postaction={decorate}] (1,0) arc[start angle=0, delta angle=180, radius=1];
        \draw[thick,decoration={markings,mark=at position 0.5 with {\arrow{>}}},postaction={decorate}] (-1,0) arc[start angle=180, delta angle=180, radius=1];
        \node at (1,0) [anchor=west,outer sep=0.1cm] {$Sp$};
        \node at (-1,0) [anchor=east,outer sep=0.1cm] {$p$};
        \node at (0,1) [anchor=south,outer sep=0.1cm] {$c$};
        \node at (0,-1) [anchor=north,outer sep=0.1cm] {$Sc$};
    \end{tikzpicture}
\]
meaning that the differential acts as $\partial c = (1 - S)p$, and it is understood that the reflection $R$ acts as $c \mapsto -Sc$, $p \mapsto p$. For the remaining $5$ circles and $E\bZ_2$ we take the cell structures we used before. Then the quotient by $\bZ_2^3$ yields a cell structure with cells
\[
    a_n c^k \; : \; n \geq 0, k \in \{0,1\}^7
\]
and nonzero differentials
\begin{align*}
    \partial c_1 &=
    \begin{cases}
        1 - \alpha\gamma & (k_2 = 0) \\
        1 - \beta\gamma & (k_2 = 1)
    \end{cases} \\
    \partial c_2 &= 1 - \alpha\beta \\
    \partial a_n &=
    \left(
        1 + (-1)^n
        \begin{cases}
            \alpha & (k_1 = 0, k_2 = 0) \\
            \beta & (k_1 = 0, k_2 = 1) \\
            \gamma & (k_1 = 1)
        \end{cases}
    \right) a_{n-1}
\end{align*}
where $\alpha=(-1)^{k_1+k_2+k_3+k_4}$, $\beta=(-1)^{k_1+k_2+k_5+k_6}$, $\gamma=(-1)^{k_1+k_3+k_5+k_7}$, and we again suppress zero-cells from the notation.

The above complexes are simple enough to analyse by hand. In particular their first and second integral homology groups, with explicit generators, are
\begin{equation} \label{eq:bases}
    \begin{split}
        H_1((T^6 \times E\bZ_2^2) / \bZ_2^2; \bZ) &= \bZ_2^8 \{ c_1, \dots, c_6, a_1, b_1 \} \,,\\
        H_2((T^6 \times E\bZ_2^2) / \bZ_2^2; \bZ) &= \bZ^3 \{ c_1 c_2, c_3 c_4, c_5 c_6 \} \\
        &\times
        \begin{aligned}[t]
            \bZ_2^{19} \{ &c_i c_j : 1 \leq i < j \leq 6, (i,j) \neq (1,2),(3,4),(5,6), \\
            & a_1 c_{1,2}, b_1 c_{3,4}, (a_1 - b_1) c_{5,6}, a_1 b_1 \}\,,
        \end{aligned} \\
        H_1((T^7 \times E\bZ_2) / \bZ_2^3; \bZ) &= \bZ_2^8 \{ c_1, \dots, c_7, a_1 \} \,,\\
        H_2((T^7 \times E\bZ_2) / \bZ_2^3; \bZ) &=
        \begin{aligned}[t]
            \bZ_2^{21} \{ &c_i c_j : 3 \leq i < j \leq 7, \\
            &c_1 c_{3,6}, c_2 c_7, a_1 c_{5,6,7}, (c_2 - a_1) c_{3,4}, (c_1 - c_2) c_{4,5}, (c_1 - a_1) c_2 \}\,.
        \end{aligned}
    \end{split}
\end{equation}
Applying $\Hom(-, U(1))$ then gives the cohomology groups \eqref{eq:t6h} and \eqref{eq:t7h} asserted in the main text. The calculations of \eqref{eq:t6t2h}, \eqref{eq:t7t3ah}, \eqref{eq:t7t3yh} follow an identical structure to these two cases.

To compute the relevant maps between these groups, we will also need to know the cup product structure. For this, one needs diagonal approximations. The standard diagonal approximation for $E\bZ_2$ is $\Delta a_n = \sum_{n_1+n_2=n} (-1)^{n_1 n_2} (a_{n_1} \times t^{n_2} a_{n_2})$ \cite{brown2012coh}, while the diagonal approximation for the first variant of the $S^1$ cell complex is $\Delta p = p p, \Delta c = cp + pc$, while for the second variant it is $\Delta p = p p, \Delta c = c(Sp) + pc$. From this one can deduce that the diagonal approximations for our two Borel quotients are
\begin{equation} \label{eq:cup}
    \begin{split}
        \Delta(a_n b_m c^k) \text{ mod } 2 &= \sum_{\substack{n_1 + n_2 = n \\ m_1 + m_2 = m}}
        a_{n_1} a_{n_2} b_{m_1} b_{m_2} \prod_{i=1}^6
        \begin{cases}
            p_i p_i & (k_i = 0) \\
            c_i p_i + p_i c_i & (k_i = 1)
        \end{cases} \\
        \Delta(a_n c^k) \text{ mod } 2 &= \sum_{n_1 + n_2 = n} a_{n_1} a_{n_2} \prod_{i=1}^7
        \begin{cases}
            p_i p_i & (k_i = 0) \\
            c_i p_i + p_i c_i & (k_i = 1)
        \end{cases}
    \end{split}
\end{equation}
at least modulo $2$, which will suffice for our purposes. Thus the mod-2 cohomology rings are
\begin{align*}
    H^*((T^6 \times E\bZ_2^2) / \bZ_2^2; \bZ_2) &= \bZ_2[\omega_1, \dots, \omega_6, \omega_a, \omega_b] / (\omega_1^2, \dots, \omega_6^2) \,,\\
    H^*((T^7 \times E\bZ_2) / \bZ_2^3; \bZ_2) &= \bZ_2[\omega_1, \dots, \omega_7, \omega_a] / (\omega_1^2, \dots, \omega_7^2)
\end{align*}
in terms of the classes which are dual to the 1-cells $\{c_1, \dots, c_6, a_1, b_1\}$ and $\{c_1, \dots, c_7, a_1\}$.

Consider now the abelianisation $\hat{G}^\ab$, which in both cases is isomorphic to $\bZ_2^8$. Let us denote the generators of its mod-2 cohomology ring as
\begin{align*}
    T^6 / \bZ_2^2: \quad H^*(B\bZ_2^8; \bZ_2) &= \bZ_2[\omega_{S_1}, \dots, \omega_{S_6}, \omega_\alpha, \omega_\beta] \,,\\
    T^7 / \bZ_2^3: \quad H^*(B\bZ_2^8; \bZ_2) &= \bZ_2[\omega_{S_3}, \dots, \omega_{S_7}, \omega_\alpha, \omega_\beta, \omega_\gamma]
\end{align*}
Then under $\hat{G} \to \hat{G}^\ab$, these classes pull back to classes on the Borel quotient. For the first case, the pullback map is trivial, and given by $\omega_{S_i} \to \omega_i,$ $ \omega_\alpha \to \omega_a, $ $\omega_\beta \to \omega_b$. To deduce the pullback map in the second case, we consider the map $\hat{G} \to \hat{G}^\ab$ at the 1-chain level. The 1-chain $c_1$ has endpoints differing by the action of $\alpha\gamma$, so maps to $\alpha\gamma$. The 1-chain $c_2$ has endpoints differing by the action of $\alpha\beta$, so maps to $\alpha\beta$. The 1-chain $a_1$ has endpoints differing by the action of $\alpha$, so maps to $\alpha$. The pullback map on cohomology is then the dual of this, namely $\omega_\alpha \to \omega_1 + \omega_2 + \omega_a, $ $\omega_\beta \to \omega_2,$ $ \omega_\gamma \to \omega_1$.

We can now study the map $H^2(B\bZ_2^8; U(1)) \to H^2(B\hat{G}; U(1))$, by taking a class on the left hand side, lifting it to $\bZ_2$ coefficients (which is always possible), applying the above pullback map, then evaluating it on the explicit basis \eqref{eq:bases} using \eqref{eq:cup}. This exercise yields the following results:
\begin{itemize}

\item For $T^6 / \bZ_2^2$, the classes $(\delta_{i,3456} \omega_\alpha + \delta_{i,1256} \omega_\beta) \omega_{S_i}$ for $1 \leq i \leq 6$ are in the kernel, the classes $\omega_{S_1} \omega_{S_2}, \omega_{S_3} \omega_{S_4}, \omega_{S_5} \omega_{S_6}$ map to the $B$-fields $\pi dx^1 \wedge dx^2, \pi dx^3 \wedge dx^4, \pi dx^5 \wedge dx^6$ in the $U(1)^3$ factor, while the remaining $28 - 6 - 3 = 19$ classes map to the $\bZ_2^{19}$ factor.

\item For $T^7 / \bZ_2^3$, the classes $(\delta_{i,1234} \omega_\alpha + \delta_{i,1256} \omega_\beta + \delta_{i,1357} \omega_\gamma) \omega_{S_i}$ for $3 \leq i \leq 7$, $(\omega_\alpha + \omega_\beta) \omega_\gamma$, and $\omega_\alpha \omega_\beta$ are in the kernel, while the remaining $28 - 7 = 21$ classes map to the $\bZ_2^{21}$ factor.

\end{itemize}
These imply \eqref{eq:19pairs} and \eqref{eq:21pairs}. Identical considerations (namely computing the action in degree 1 then deducing the action on degree 2 via the cup product) apply for the other pullback maps.

\subsection{Via group extensions}

In the main text, we noted that in our case $H^p_G(T^n;U(1))=H^p(B\hat{G};U(1))$
where $\hat{G}$ sits in a short exact sequence $1\to \bZ^n\to \hat G\to G\to 1$.
We can then determine $H^2(B\hat G;U(1))$ using the standard fact that 
it classifies extensions $1\to U(1)\to \tilde G\to \hat G\to 1$,
by directly studying all possible $\tilde G$.

\subsubsection{$T^6/\bZ_2^2$}

The structure of $\hat G$ was given in \cref{foot:T6hatG}.
We fix lifts $V_{\alpha,\beta}$ of $\alpha,\beta \in \hat G$ to $\tilde G$ 
via the condition $V_\alpha^2=V_\beta^2=1$. 
We also pick lifts $V_i$ of $T_i\in \hat G$ to $\tilde G$.
Then any element of $\tilde G$ is given by an element of $U(1)$ times \begin{equation}
V_\alpha^{n_\alpha} V_\beta^{n_\beta} V_1^{n_1} V_2^{n_2} V_3^{n_3} V_4^{n_4} V_5^{n_5} V_6^{n_6},\label{canonical6}
\end{equation}
where $n_{\alpha,\beta}=0,1$ and $n_i\in \bZ$.
To fix the structure of $\tilde G$, we then need to be able to multiply 
two elements of the form \eqref{canonical6} and put it back to the same form.
For this, we need to know the phases $c_{ij}$, $a_i$, $b_i$ and $q$ 
appearing in the commutation relations \begin{align}
V_i V_j &= c_{ij} V_j V_i, \qquad (i<j) \\
V_\alpha V_i &= a_i V_i^{\alpha_i } V_\alpha, \\
V_\beta V_i &= b_i V_i^{\beta_i} V_\beta, \\
V_\alpha V_\beta &= q V_\beta V_\alpha,
\end{align}
where $\alpha_i$ and $\beta_i$ are signs $\pm1$ so that $\alpha T_i = T_i^{\alpha_i} \alpha$
and $\beta T_i = T_i^{\beta_i} \beta$.

Using $V_\alpha^2=V_\beta^2=1$, we can show that $a_i^2=b_i^2=q^2=1$.
When $\alpha_i=-1$, we have $V_\alpha V_i = a_i V_i^{-1} V_\alpha$
with $a_i=\pm1$.
We can change this sign $a_i$ by redefining $V_i^\text{new} \coloneqq \sqrt{-1}V_i$.
Similarly, when $\beta_i=-1$, we have $V_\beta V_i = b_i V_i^{-1} V_\beta$
with $a_i=\pm1$.
We can change this sign $b_i$ by redefining $V_i^\text{new} = \sqrt{-1}V_i$ as before.
But if $\alpha_i=\beta_i=-1$, the relative sign $a_ib_i$ cannot be removed,
which appears in the commutation relation $V_\alpha V_\beta V_i = a_i b_i V_\alpha V_\beta$.
These  account for the seven phases \begin{equation}
\braket{\alpha, S_{1,2}}, \braket{\beta, S_{3,4}}, \braket{\alpha\beta, S_{5,6}},
\braket{\alpha, \beta}
\end{equation} of \eqref{eq:19pairs}.

We can also show, by considering $V_\alpha V_i V_j$
that $c_{ij}^2=1$ if $\alpha_i \alpha_j =-1$.
Similarly, $c_{ij}^2=1$ if $\beta_i\beta_j=-1$.
Otherwise $c_{ij}$ are unconstrained. 
We find that $c_{ij}$ for $(i,j)=(1,2)$, $(3,4)$, $(5,6)$ are unconstrained
while other $c_{ij}$ are $\pm1$, accounting for the other twelve $\pm1$ phases 
$\braket{S_i,S_j}$ of \eqref{eq:19pairs}.

\subsubsection{$T^7/\bZ_3^2$}

The structure of $\hat G$ was given in \cref{foot:T7hatG}.
We fix lifts $V_{\alpha,\beta,\gamma}\in \tilde G$ of $\alpha,\beta,\gamma\in \hat G$,
using $V_{\alpha,\beta,\gamma}^2=1$.
We fix lifts $V_i\in \tilde G$  of $T_i\in \hat G$ for $i=3,4,5,6,7$.
We declare the lifts $V_{1,2}\in\tilde G$ of $T_{1,2}\in \hat G$  to be \begin{equation}
V_1 \coloneqq V_\beta V_\alpha V_\beta V_\alpha,\qquad
V_2 \coloneqq V_\gamma V_\alpha V_\gamma V_\alpha.
\label{V12}
\end{equation}
Then any element of $\tilde G$ is an element of $U(1)$ multiplied by
an element of the form \begin{equation}
V_\alpha^{n_\alpha} V_\beta^{n_\beta} V_\gamma^{n_\gamma} V_1^{n_1} V_2^{n_2} V_3^{n_3} V_4^{n_4} V_5^{n_5} V_6^{n_6} V_7^{n_7},\label{canonical7}
\end{equation}
where $n_{\alpha,\beta,\gamma}=0,1$ and $n_i\in \bZ$.

To fix the structure of $\tilde G$, we then need to be able to multiply 
two elements of the form \eqref{canonical7} and put it back to the same form.
For this, we need to know the phases $c_{ij}$, $a_i$, $b_i$, $c_i$ and $q$ 
appearing in the commutation relations \begin{align}
V_i V_j &=  c_{ij} V_j V_i, \qquad (i<j)\\
V_\alpha V_i &= a_i V_i^{\alpha_i } V_\alpha, \\
V_\beta V_i &= b_i V_i ^{\beta_i} V_\beta, \\
V_\gamma V_i &= c_i V_i ^{\gamma_i} V_\beta, \\
V_\beta V_\gamma &= q V_\gamma V_\beta V_1^{-1}.
\end{align}
Here, $i,j=3,4,5,6,7$, and
$\alpha_i$, $\beta_i$ and $\gamma_i$ are signs $\pm1$ so that
$\alpha T_i = T_i^{\alpha_i} \alpha$,
$\beta T_i = T_i^{\beta_i} \beta$,
$\gamma T_i = T_i^{\gamma_i} \gamma$.
Note that there is no need to explicitly introduce undetermined phases 
in the commutation relation between $V_{\alpha}$ and $V_{\beta}$
or that between $V_{\alpha}$ and $V_{\gamma}$,
since they can be absorbed into the definitions of $V_1$ and $V_2$  in \eqref{V12}.

By the same tricks in the case of $T^6/\bZ_2^2$,
we can show that $c_{ij}$, $a_i$, $b_i$, $c_i$ and $q$ are all $\pm1$.
Among them, $c_{ij}$ provide 10 signs $\braket{S_i,S_j}$,
in the first line of \eqref{eq:t7t3yw}.
Furthermore, just as before, redefining $V_i$ by $V_i^\text{new} \coloneqq \sqrt{-1}V_i$,
we can flip $a_i$ if $\alpha_i=-1$, and so on.
We see that invariant phases among $a_i$, $b_i$ and $c_i$ are
$a_{5,6,7}$, $b_{3,4,7}$, $c_{4,6}$, $a_3c_3$ and $b_5c_5$.
They provide 5 signs 
$\braket{\alpha, S_{5,6,7}}$, $\braket{\beta, S_{3,4,7}}$, 
$\braket{\gamma, S_{4,6}}$, $\braket{\alpha\gamma, S_3}$, 
$\braket{\beta\gamma, S_5}$ in the second line of \eqref{eq:t7t3yw}.
Finally, $q$ appears as the commutation relation \begin{equation}
(V_\alpha V_\beta) V_\gamma = q V_\gamma(V_\alpha V_\beta),
\end{equation}
providing the last sign $\braket{\alpha\beta,\gamma}$ in the last line of \eqref{eq:t7t3yw}.

\bibliographystyle{ytamsalpha}
\def\arxivfont{\rm}
\bibliography{ref}

\end{document}